\documentclass[sigconf, nonacm]{acmart} 
\usepackage{package} 
\usepackage{def} 
\newcommand\vldbdoi{10.14778/3551793.3551828}
\newcommand\vldbpages{XXX-XXX}
\newcommand\vldbvolume{XX}
\newcommand\vldbissue{XX}
\newcommand\vldbyear{XXXX}
\newcommand\vldbauthors{\authors}
\newcommand\vldbtitle{\shorttitle} 
\newcommand\vldbavailabilityurl{https://github.com/msr-fiddle/harmony} 
\newcommand\vldbpagestyle{empty} 
\begin{document}
\title{\sysname: Overcoming the Hurdles of GPU Memory Capacity to Train Massive DNN Models on Commodity Servers} 
\author{Youjie Li}
\affiliation{%
  \institution{UIUC}
}
\email{li238@illinois.edu}
\authornote{Work done as a Project Fiddle intern at MSR.}

\author{Amar Phanishayee}
\affiliation{%
  \institution{Microsoft Research}
}
\email{amar@microsoft.com}

\author{Derek Murray}
\affiliation{%
  \institution{Lacework}
}
\email{derek.murray@lacework.net}
\authornote{Work done when the author was at Microsoft.}

\author{Jakub Tarnawski}
\affiliation{%
  \institution{Microsoft Research}
}
\email{jakub.tarnawski@microsoft.com}

\author{Nam Sung Kim}
\affiliation{%
  \institution{UIUC}
}
\email{nam.sung.kim@gmail.com}

\begin{abstract}

Deep neural networks (DNNs) have grown exponentially in size over the past decade, leaving only those who have massive datacenter-based resources with the ability to develop and train such models. One of the main challenges for the long tail of researchers who might have only limited resources (e.g., a single multi-GPU server) is limited GPU memory capacity compared to model size. The problem is so acute that the memory requirement of training massive DNN models can often exceed the aggregate capacity of all available GPUs on a single server; this problem only gets worse with the trend of ever-growing model sizes. Current solutions that rely on virtualizing GPU memory (by swapping to/from CPU memory) incur excessive swapping overhead. 
In this paper, we present a new training framework, \sysname{}, and advocate rethinking how DNN frameworks schedule computation and move data to push the boundaries of training massive models efficiently on a single commodity server. Across various massive DNN models, \sysname is able to reduce swap load by up to two orders of magnitude and obtain a training throughput speedup of up to 7.6$\times$ over highly optimized baselines with virtualized memory.

\end{abstract}

\maketitle
\pagestyle{\vldbpagestyle}
\begingroup\small\noindent\raggedright\textbf{PVLDB Reference Format:}\\
\vldbauthors. \vldbtitle. PVLDB, \vldbvolume(\vldbissue): \vldbpages, \vldbyear.\\
\href{https://doi.org/\vldbdoi}{doi:\vldbdoi}
\endgroup
\begingroup
\renewcommand\thefootnote{}\footnote{\noindent
This work is licensed under the Creative Commons BY-NC-ND 4.0 International License. Visit \url{https://creativecommons.org/licenses/by-nc-nd/4.0/} to view a copy of this license. For any use beyond those covered by this license, obtain permission by emailing \href{mailto:info@vldb.org}{info@vldb.org}. Copyright is held by the owner/author(s). Publication rights licensed to the VLDB Endowment. \\
\raggedright Proceedings of the VLDB Endowment, Vol. \vldbvolume, No. \vldbissue\ %
ISSN 2150-8097. \\
\href{https://doi.org/\vldbdoi}{doi:\vldbdoi} \\
}\addtocounter{footnote}{-1}\endgroup

\vspace{-0.5ex}
\ifdefempty{\vldbavailabilityurl}{}{
\vspace{.3cm}
\begingroup\small\noindent\raggedright\textbf{PVLDB Artifact Availability:}\\
The source code, data, and/or other artifacts have been made available at \url{\vldbavailabilityurl}.
\endgroup
}

\section{Introduction}
\label{sec:intro}

Modern DNNs have transformed our approach of solving a range of problems such as image classification~\cite{krizhevsky2012imagenet}, semantic segmentation~\cite{sun2019HRNetV2}, translation~\cite{wu2016google}, and language modeling~\cite{radford2019gpt2}.
Over the years, these models have grown exponentially in size while continuing to achieve unprecedented accuracy on ever more complex tasks~\cite{kaplan2020scaling, angermueller2016deep, mnih2013playing}.
For example, a 557-million-parameter AmoebaNet can achieve super-human accuracy in image classification~\cite{huang2018gpipe}.
Similarly, a state-of-the-art language model like the 175-billion parameter GPT-3~\cite{brown2020language} can generate human-like text~\cite{ram2020gpt3,manjoo2020gpt3,guardian2020gpt3}.
Training these models to accuracy takes weeks to months of wall-clock time, despite running in parallel on large clusters of fast accelerators.

These resource demands leave only those who have massive datacenter-based resources (e.g., Google, Microsoft, NVIDIA, etc.) with the ability to train such models.  
The long tail of researchers who have only limited resources (e.g., a single server with multiple GPUs) increasingly risk being alienated from innovating in this space.  
While training on larger clusters naturally results in speedier training, in this paper we investigate how to push the boundaries of training massive models on 
\textit{a single commodity server} \--- a setting invaluable for developing, debugging, and fine-tuning DNNs~\cite{eliad2021ftpipe}.

\begin{figure}[!t]
  \centering
  \includegraphics[width=\linewidth]{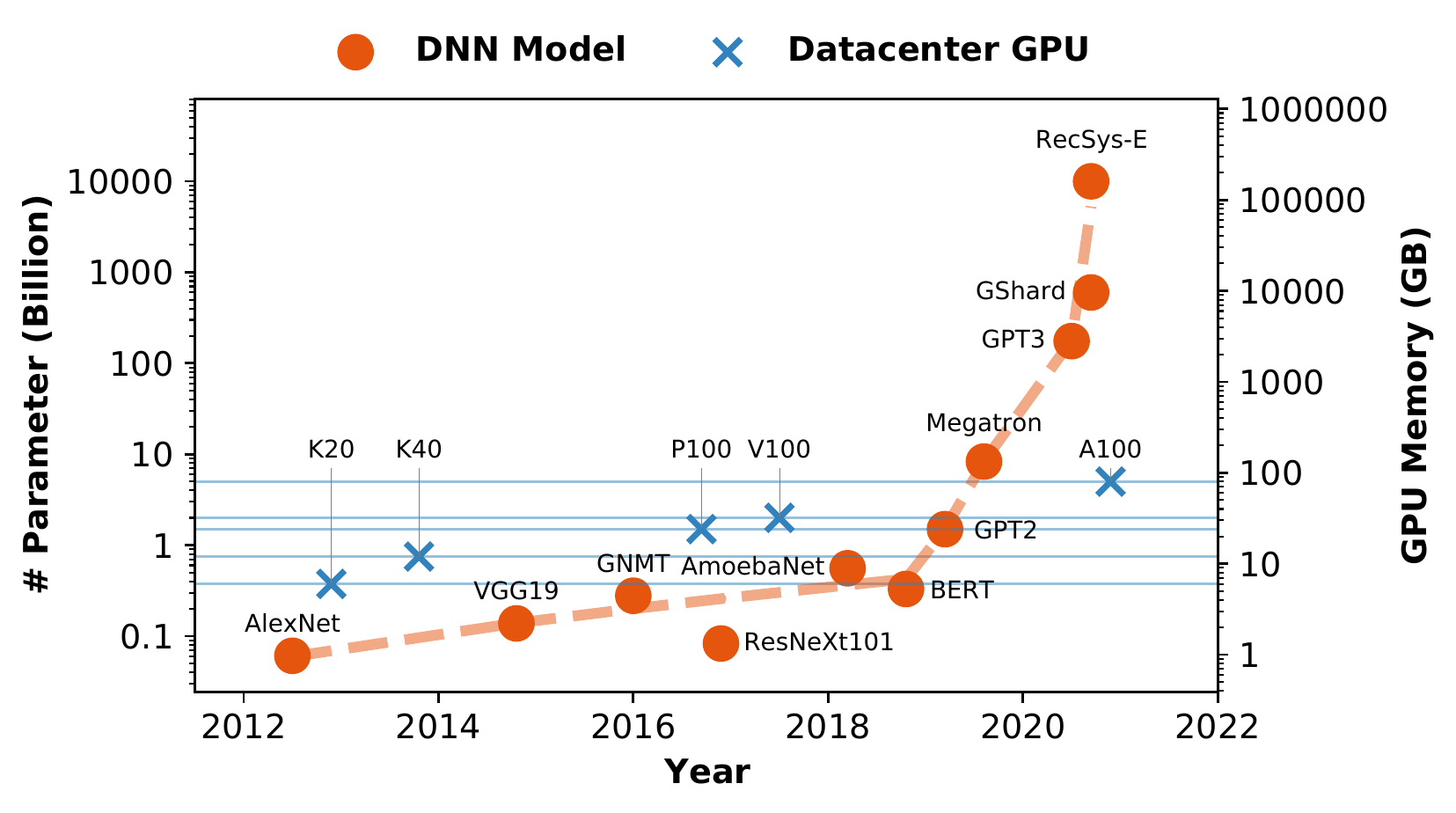}
  \vspace{-4ex}
  \caption{Growth of DNN model size and GPU memory capacity over the past decade~\cite{gholami2020ai_and_memory_wall,TeslaGPU}. 
        Memory consumed here only accounts for model state which is a small fraction of total training memory footprint~\cite{rhu2016vdnn,rajbhandari2019zero,jain2018gist,chen2016training,wang2018superneurons,gholami2020ai_and_memory_wall}.}
  \label{fig:modelsize}
  \vspace{-2ex}
\end{figure}
\begin{figure*}[!t]
  \centering
  \includegraphics[width=\linewidth]{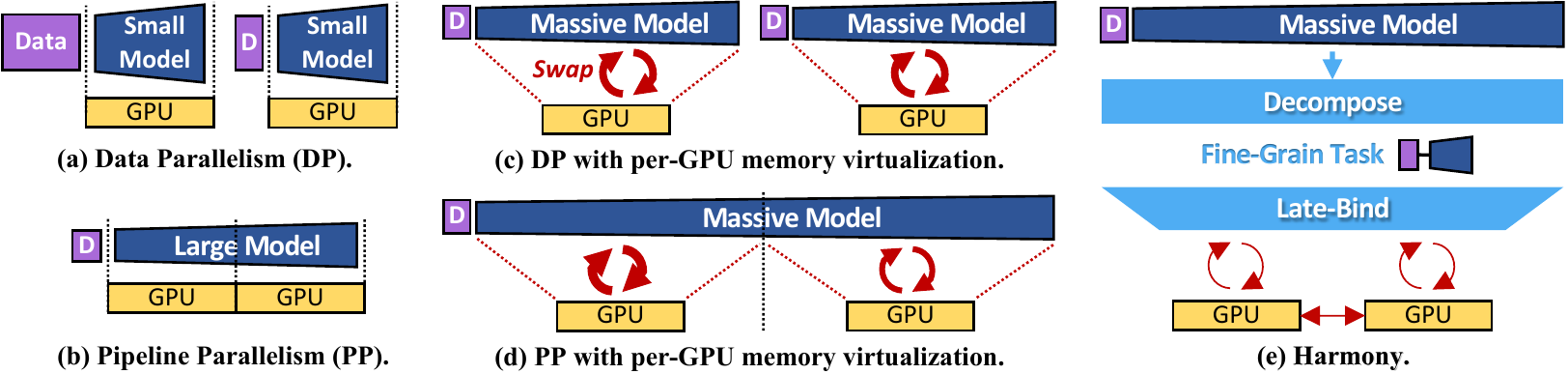}
  \vspace{-5ex}
  \caption{Illustrative comparison between different approaches for massive model training. (Only one data batch is shown.)}
  \label{fig:concept}
\end{figure*}

\vspace{1ex}
\niparagraph{Challenges.} 
One of the main challenges in training massive models is that the required memory footprint
far exceeds the memory capacity of
accelerators.
Figure~\ref{fig:modelsize} shows how sizes of image classification and language models have grown dramatically over time.
Furthermore, model parameters are only a small part of the memory footprint of training; 
gradients, stashed activations, optimizer states, and framework workspace all taken together significantly blow up the memory footprint~\cite{rhu2016vdnn,rajbhandari2019zero,jain2018gist,chen2016training,wang2018superneurons}.


This memory footprint problem motivates recent innovations that alleviate memory pressure.
For example, recent advances in GPU memory virtualization push the boundaries of what can be achieved on a single GPU~\cite{rhu2016vdnn,peng2020capuchin,huang2020swapadvisor,cho2018lms}, but as we show in \secref{sec:motivation} such techniques are inefficient when applied to parallel multi-GPU training regardless of data parallelism~\cite{li2014scaling,DDP} or pipeline parallelism~\cite{huang2018gpipe,narayanan2019pipedream}, as illustrated in Figure~\ref{fig:concept}(a--d).
Other techniques, such as encoding data structures~\cite{jain2018gist}, recomputing intermediate tensors~\cite{chen2016training}, 
sharding optimizer~\cite{rajbhandari2019zero}, 
offloading optimizer to CPU~\cite{ren2021zerooffload}, 
and splitting a large layer into small ones~\cite{shoeybi2019megatron},
all aim to reduce memory pressure during training.
However, despite these optimizations, the general problem of \emp{efficiently training massive models on a single server with a handful of commodity GPUs while exhausting the collective memory capacity of all available GPUs and CPU DRAM}
is still an open problem.

\begin{figure*}[!t]
\centering
    \begin{subfigure}{.33\textwidth}
      \centering
      \includegraphics[width=0.98\linewidth]{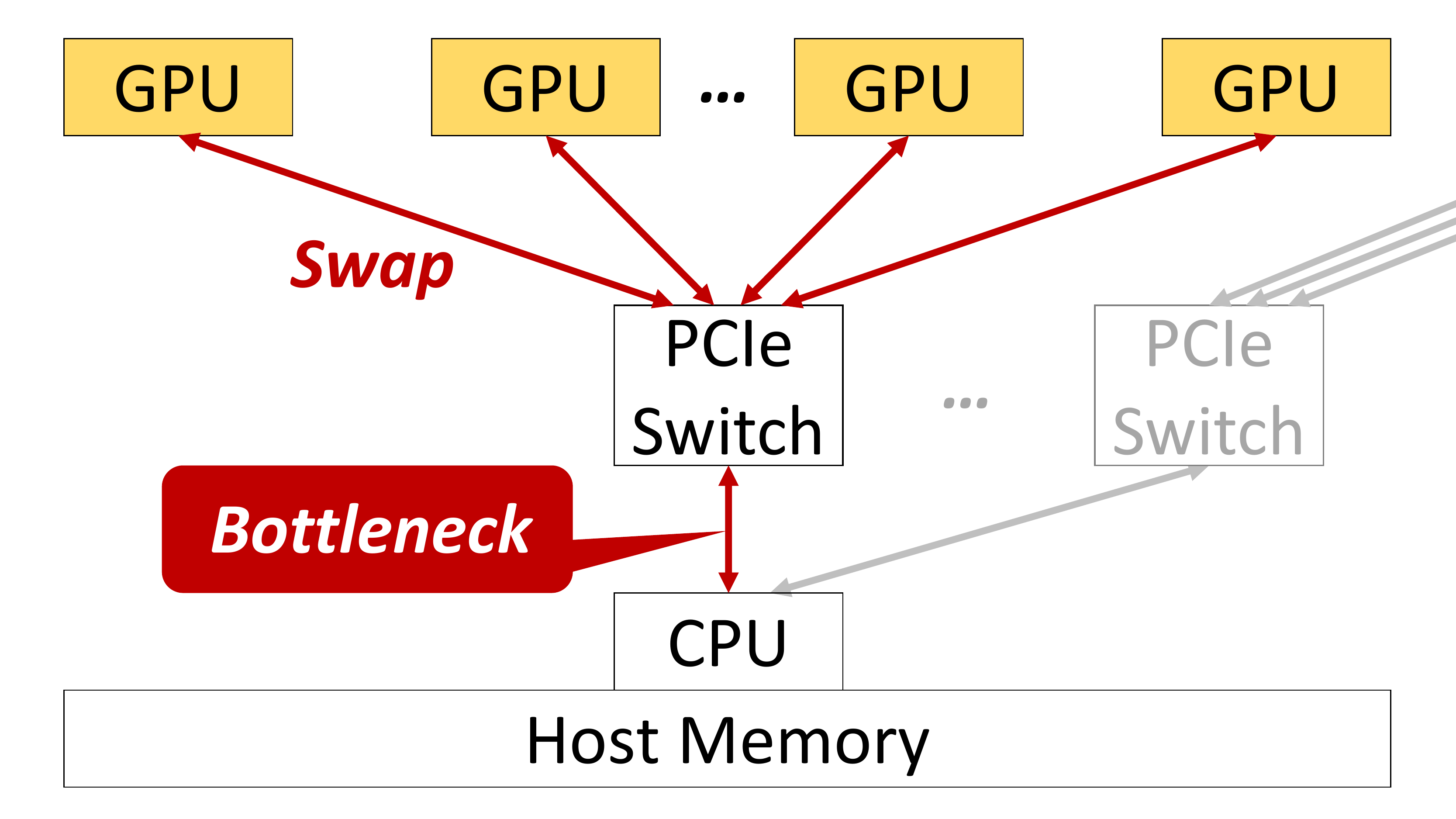}
      \vspace{-1ex}
      \caption{Intra-server interconnects.} 
      \label{fig:pcie}
      \vspace{-2ex}
    \end{subfigure}
    \begin{subfigure}{.33\textwidth}
      \flushleft
      \includegraphics[width=0.98\linewidth]{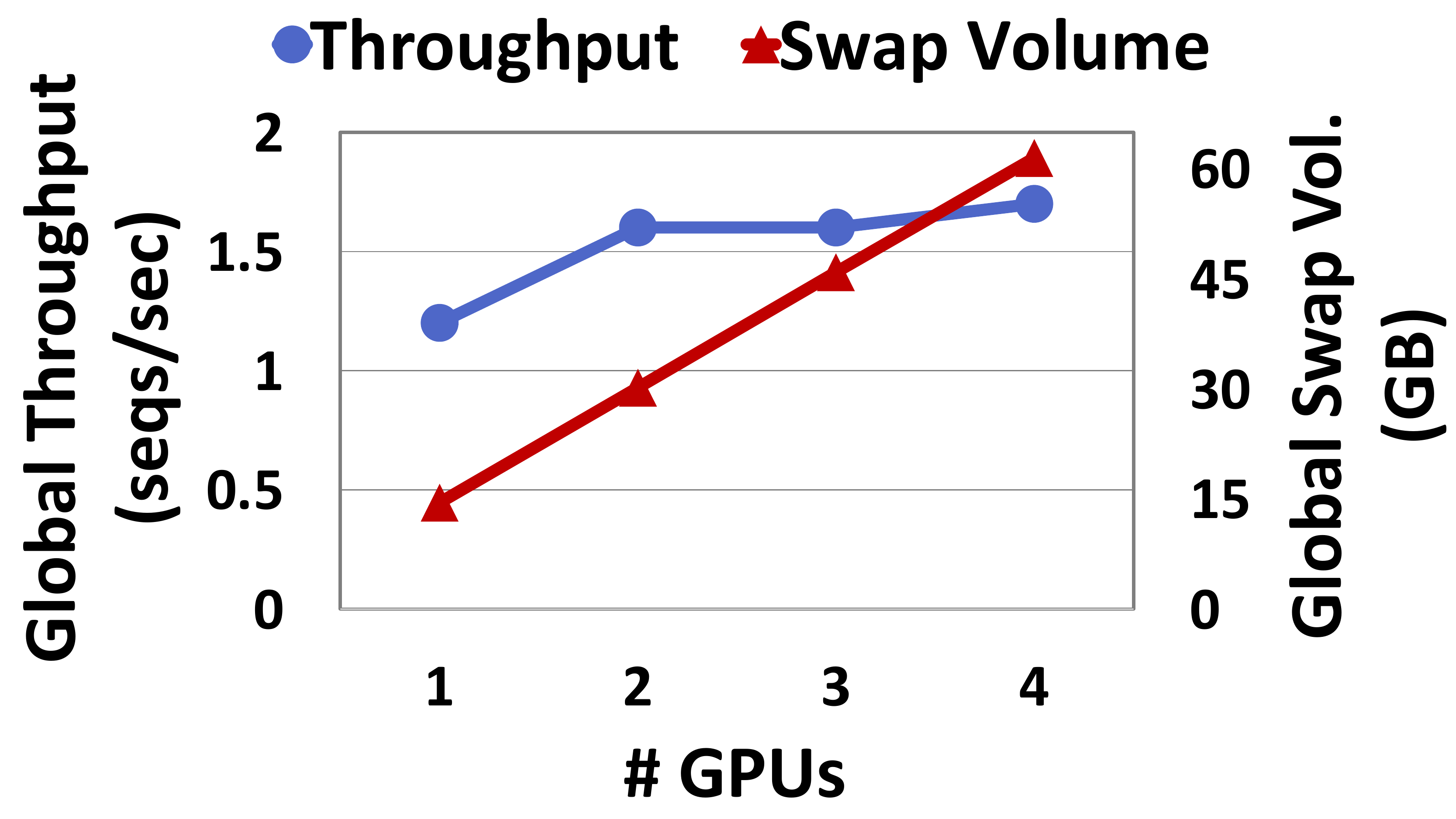}
      \vspace{-1ex}
      \caption{DP with per-GPU memory virtualization.}
      \label{fig:dp_tput_swap}
      \vspace{-2ex}
    \end{subfigure}
    \begin{subfigure}{.33\textwidth}
      \flushright
      \includegraphics[width=0.98\linewidth]{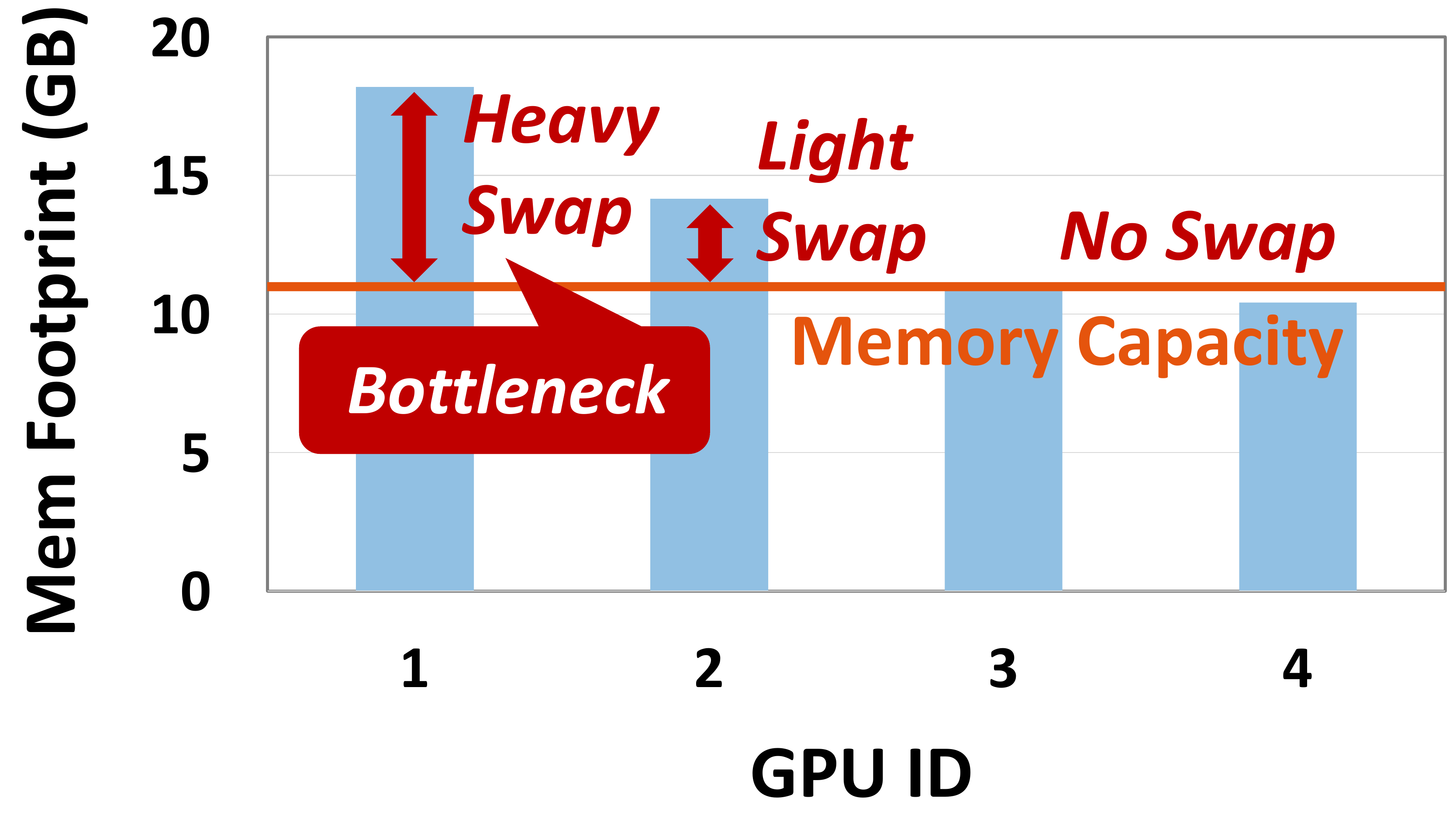}
      \vspace{-1ex}
      \caption{PP with per-GPU memory virtualization.}
      \label{fig:pp_swap}
      \vspace{-2ex}
    \end{subfigure}
\caption{The CPU-GPU swap bottleneck in Data Parallelism (DP)~\cite{li2020pytorch} and Pipeline Parallelism (PP)~\cite{narayanan2020memory} when using GPU memory virtualization.
For example, training BERT~\cite{devlin2018bert} on a server with four GTX1080Tis (11GB) and a batch size of 5 results in memory footprint exceeding GPU memory capacity, requiring IBM-LMS~\cite{PTLMS} for virtualizing individual GPU memory.
(a)~and~(b) show that DP's swap volume increases linearly with the number of GPUs, exposing the bottleneck PCIe link and thus throttling training throughput. 
(c) shows that PP's swap volume is unbalanced across GPUs, resulting in pipeline bottleneck.}
\label{fig:swap_bottleneck}
\end{figure*}

We argue that current DNN frameworks have two fundamental problems that limit massive model training on modest deployments.
First, they \emp{schedule work at a coarse granularity}, treating the training program as a black box: computing an entire model or an entire stage of layers for each input batch.
This coarse granularity limits flexibility of scheduling tasks to available resources, thus thwarting memory-reuse--based performance enhancements that can reduce virtual memory swap overhead.
For example, executing a group of DNN layers (even with intra-layer partitions), one input batch at a time, limits reuse of weights loaded into memory across different input batches, as they might get swapped out.
Second, frameworks \emp{eagerly bind work to accelerators}, pushing this decision all the way to the programmer's training script in most cases.
For example, in PyTorch~\cite{li2020ddp}, the state of a layer group is bound to a user-script--defined device, and thus the forward and backward computation on that state is implicitly bound to the same device. 
Virtualizing the memory of a single GPU helps here, by treating the nearby host RAM as a swap target, but it makes inefficient use of other available GPUs and the interconnects between them.

\vspace{1ex}
\niparagraph{Contributions.}
Ideally, users could write DNN training programs that target a single virtual accelerator with practically unbounded memory. 
Our proposed system, \sysname, targets this ideal.
As illustrated in Figure~\ref{fig:concept}(e), \sysname decomposes a model's operations in a training script into \emp{fine-grained tasks}~\footnote{A task consists of an input microbatch and a contiguous set of layers; there is no requirement of one-one correspondence between forward and backward tasks.} 
and introduces a novel task scheduler that efficiently maps computation and state to physical devices (\emp{late binding}); 
the tasks in the task graph can run on different physical devices in a data- or pipeline-parallel fashion and \system{} transparently moves state and data across tasks.
Unlike prior pipeline-parallel training~\cite{huang2018gpipe,narayanan2019pipedream,narayanan2020memory}, 
each GPU in \sysname{} no longer hosts a fixed stage of layers, thus resulting in a novel pipelining scheme, \textbf{\textit{\wap}}, while offering synchronous SGD semantics.

\sysname{} has to overcome two main challenges to operate at peak throughput: (i) \textit{minimizing expensive CPU-GPU memory swaps}, and (ii) \textit{balancing load across all GPUs} so that there is no bottleneck worker in the execution pipeline.
\sysname{} achieves this by using four distinct optimizations for efficient training:

\niparagraph{\circled{1} \textit{Reusing State in GPU Memory across Different Inputs}}. 
 Empowered by the flexibility of scheduling at a finer granularity, we propose a new technique called \textit{input-batch grouping}, where a scheduled layer(s) can run across a group of input batches before scheduling the next layer(s) on the same GPU, thus improving state reuse in GPU memory and consequently improving arithmetic intensity. 
 
\niparagraph{\circled{2} \textit{Scheduling Tasks Just-in-time}}. 
 \system{} schedules tasks as soon as all input dependencies are available, thus avoiding the risk of swapping out those dependencies; this especially helps tasks such as weight update, which in frameworks such as PyTorch are normally scheduled to execute only after the backward pass for the entire model, resulting in avoidable CPU-GPU swaps.
 
\niparagraph{\circled{3} \textit{Generalized Tensor Swaps over Fast Peer-to-peer Links}}.
 With \textit{late binding} of tasks to GPUs, \sysname{} places adjacent tasks across GPUs and swaps tensors directly between GPUs using \textit{peer-to-peer (p2p) swaps} rather than swapping state back and forth to CPU memory.
 Unlike prior work, p2p swaps in \sysname{} are not limited to only the output tensors of stages~\cite{huang2018gpipe,fan2021dapple,narayanan2019pipedream} but can be used to transfer or swap any intermediate tensor within each stage.
 
\niparagraph{\circled{4} \textit{Multi-dimensional Layer Packing}}.
 Tensor swaps can be minimized by packing contiguous layers together. 
 Greedily picking the largest pack size that fits a GPU, however, results in globally sub-optimal pipelines due to imbalance across GPUs.
 Furthermore, picking layer packs is challenging because not all layers are created equal. 
 The same layer has drastically different compute and memory requirements between forward and backward passes for a fixed batch size; the differences are only accentuated when we consider different batch sizes.
 We thus have to find packs in the multi-dimensional space (forward batch size, forward packs, backward batch size, backward packs) that balance compute, memory, and swaps across GPUs; however, we find this problem of \textit{optimally determining layer packs in \sysname{} to be NP-hard}.
 We propose an efficient heuristic algorithm that searches through this multi-dimensional space to find effective parallel training schedules without pipeline bottlenecks. 
 \emp{To our best knowledge, no prior work has attempted such multi-dimensional layer packing.}

In this paper, we show how \textit{task decomposition} and \textit{late binding}, together with a set of novel performance optimizations mentioned above, 
enable virtualized parallel training of massive DNNs that exhaust collective memory capacity of all available accelerators in modest single-server deployments\footnote{We omit storage from the memory hierarchy; if incorporated, our work can target even larger models that exceed CPU DRAM capacity.}.
A short workshop version of this paper highlighted the limitations of existing DNN frameworks in training massive models~\cite{li2021harmony}.
Unlike our prior work, in the current paper we provide foundational principles, a detailed design backed by a concrete implementation, and extensive evaluations across various massive models.
We show that \sysname is able to \emp{reduce swap load by up to two orders of magnitude} and obtain 
a training throughput \emp{speedup of up to 7.6$\times$} over highly optimized baselines with virtualized memory, including recent systems such as \zinf~\cite{rajbhandari2021zeroinf}, while offering \emp{synchronous SGD semantics}.

\vspace{1ex}
\niparagraph{Roadmap.}
In the rest of this paper, we first present the limitations of related works with a focus on GPU memory virtualization (\secref{sec:motivation}), then offer a high-level overview of \sysname{}
(\secref{sec:approach}), followed by low-level designs and implementations (\secref{sec:impl}). 
We experimentally validate \sysname{}'s efficacy (\secref{sec:eval}) before concluding (\secref{sec:discussion}-\ref{sec:conclusion}).

\section{Background and Related Works}
\label{sec:motivation}

\niparagraph{Parallel Training.}
Data Parallelism (DP)~\cite{li2014scaling,li2020ddp,li2018pipe}, the predominant mode of parallel DNN training, requires the entire model's memory footprint to fit on each GPU, making it unfit for massive model training (Figures~\ref{fig:modelsize} and~\ref{fig:concept}(a)). 
Pipeline Parallelism (PP)~\cite{narayanan2019pipedream,narayanan2020memory,huang2018gpipe,fan2021dapple}  and Model Parallelism (MP)~\cite{shoeybi2019megatron} have become mainstream for training large models by partitioning a model so that each part fits on an individual GPU (Figure~\ref{fig:concept}(b)).
However, even in the face of partitioned models, all these systems \textit{require training memory footprint to be less than the collective memory capacity of all GPUs}.


\niparagraph{Memory Optimizations.}
To reduce the memory footprint, modern frameworks incorporate various memory optimizations by default, 
such as the \textit{recompute} that re-materializes intermediate tensors when needed~\cite{chen2016training,wang2018superneurons,jain2020checkmate}.
CPU-offloading is also used for offloading model/optimizer states from GPU to CPU~\cite{PTBERT,TFBERT,ren2021zerooffload,rajbhandari2021zeroinf}.

\niparagraph{GPU Memory Virtualization.}
Despite various memory optimizations, GPU memory virtualization~\cite{UVM} remains inexorable due to the exponential growth in model sizes. 
Recent work has applied this idea to train large DNNs by backing GPU memory with CPU memory and swapping tensors between CPU and GPU~\cite{rhu2016vdnn,peng2020capuchin,huang2020swapadvisor,cho2018lms}.
However, such techniques are limited to only an \emp{individual GPU} considered in isolation.
Here we show that per-GPU memory virtualization is inefficient as it causes either a high swap overhead when used in DP or swap imbalance in PP (Figure~\ref{fig:concept}(c--d)).

Today's frameworks have four key inefficiencies that cause these swap-overhead related performance problems in parallel training:

\niparagraph{\circled{1} \textit{Repeated Swaps.}} A layer can consume different input data batches or intermediate tensors at different times, but it always requires the same weight or gradient buffer.
With GPU memory virtualization, these common weight and gradient are swapped in and out repeatedly across batches of input data. 

\niparagraph{\circled{2} \textit{Unnecessary Swaps.}} Certain operators in DNN frameworks today are scheduled at rigid points in the timeline of a training iteration, even though all their inputs are available much earlier.
When training massive models with GPU virtualization, this rigidity is inefficient: 
the GPU-resident inputs and state for such operators can be swapped out of GPU memory, only to be swapped back in again when the operator is actually scheduled.
For example, in PyTorch, the weight update for each layer only starts after the backward pass of the entire model, potentially causing unnecessary swaps of most layer weights and gradients.

\niparagraph{\circled{3} \textit{Only CPU-GPU Swaps.}} GPU memory virtualization lacks context about parallel training, works in isolation to other GPUs, and can only swap to host memory.  
This exposes the bottleneck device-to-host interconnect (Figure~\ref{fig:swap_bottleneck}(a)) and misses the opportunity to use fast device-to-device links for cross-device swaps. 
Figure~\ref{fig:swap_bottleneck}(b) shows that in DP, the swap overhead across multiple GPUs throttles throughput, as the global swap load exposes the bottleneck link: CPU and shared PCIe links with 1:4$\sim$8 over-subscription~\cite{li2019evaluating,DGX1,DGX2,ASUS,PNY}. 
As each GPU is swapping a similar amount of state, the swap overhead grows linearly in the number of GPUs.
Furthermore, PP may use p2p communication but only for per-stage output tensors (a small fraction of all tensors); it leaves all intermediate tensors within each stage, thus swapping them to the CPU when combined with per-GPU memory virtualization.

\niparagraph{\circled{4} \textit{Unbalanced Swaps.}} In PP, pipeline stages are designed to be compute-load balanced, but such pipelining inherently has imbalanced memory sizes across stages: the head of the pipeline must stash more activations compared to the tail~\cite{narayanan2019pipedream, narayanan2020memory}.
Lacking this context and operating in isolation on individual GPUs, naively virtualizing GPU memory can result in swap imbalance across stages, exposing the bottleneck stage with the heaviest swap (Figure~\ref{fig:swap_bottleneck}(c)).

\section{Training in \sysname{}}
\label{sec:approach}

\begin{figure}[!t]
  \centering
  \includegraphics[width=\linewidth]{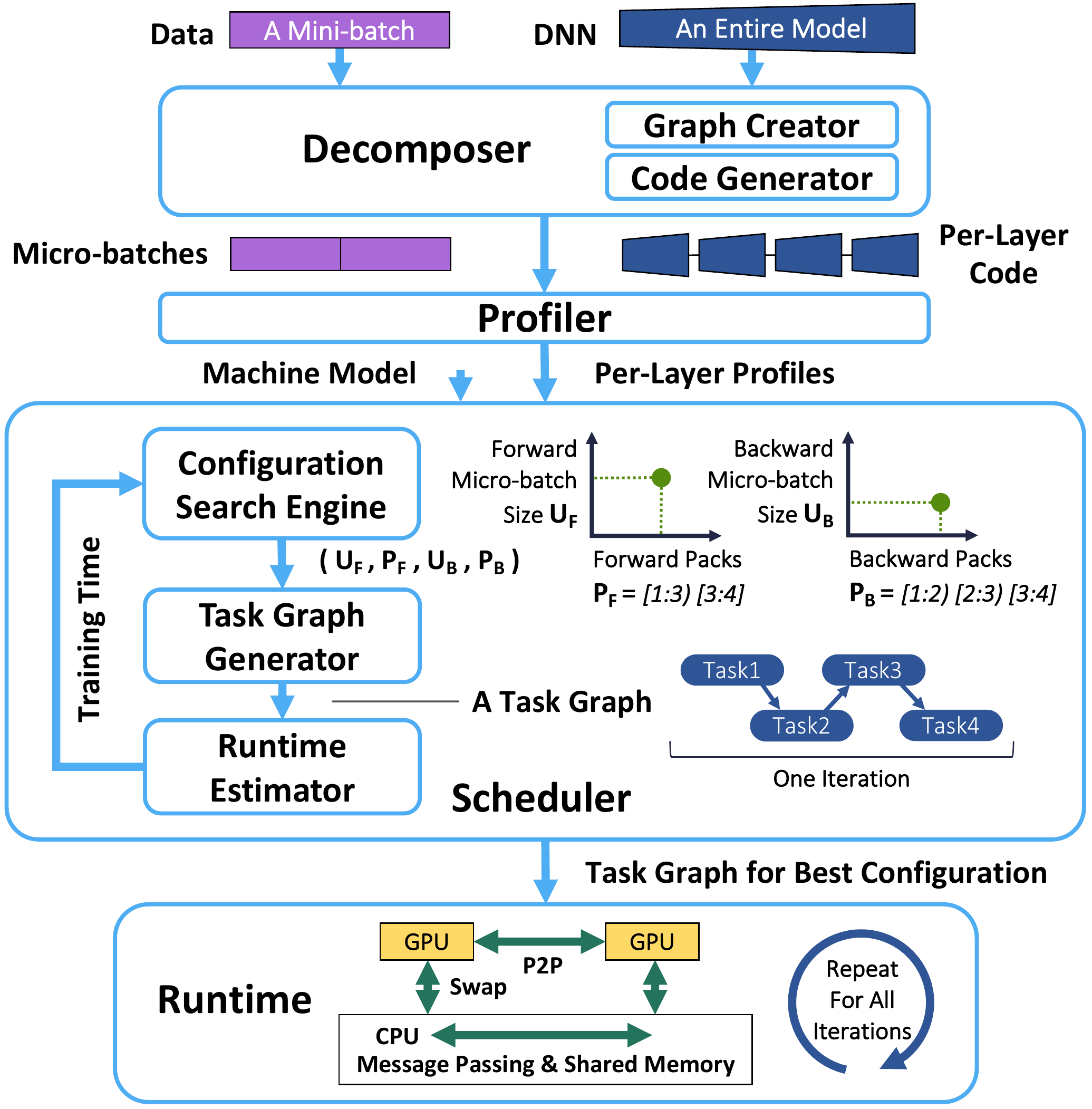}
  \vspace{-4ex}
  \caption{High-level overview of \sysname{}.}
  \label{fig:overview}
  \vspace{-2ex}
\end{figure}


Figure~\ref{fig:overview} shows a high-level overview of \sysname{}.
First, users provide \sysname{} with training data and their model (written in imperative-style PyTorch~\cite{PyTorch}, as if running sequentially on one device). \sysname{}'s \textit{Decomposer} breaks down the entire model by extracting its layer-granularity graph (via the \textit{Graph Creator}), and then generating per-layer code based on the graph such that they can be executed individually if needed (via the \textit{Code Generator}).
The data minibatch is also decomposed into small microbatches.

Next, \sysname{}'s \textit{Profiler} executes the layer-granularity graph, one layer at a time, by running the per-layer code on a single GPU of the type that will be used in the deployment (seamlessly swapping tensors between CPU and GPU as required); it does this both for the forward pass and also later for the backward pass when the graph is traversed in the reverse direction.  
The profiler repeats this process across different microbatch sizes. 
This generates profiles containing computation times, memory footprint, and input tensor sizes for each layer under different settings.

Then, \sysname{}'s \textit{Scheduler} takes the generated profiles along with the machine model (e.g., GPU memory capacity, number of GPUs, and interconnects) to compile a schedule of a \textit{single} training iteration.  It does this by: 1) selecting which layers should be executed together as a pack and thus picking a training configuration (a four-tuple of \textbf{\textit{$<$forward microbatch size $U_F$, forward layer packs $P_F$, backward microbatch size $U_B$, backward layer packs $P_B {>}$}}), 2) building a task graph for this configuration (\textit{Task Graph Generator}), 3) estimating its training time (\textit{Runtime Estimator}), and 4) refining the configuration by searching through the space of configuration options (\textit{Configuration Search Engine}).

Finally, once the best configuration is found and the final task graph is generated, the \sysname{} \textit{Runtime} then executes it for all training iterations on the set of GPUs in the deployment.

\niparagraph{Modes of Parallel Execution.} \sysname{} supports two modes of execution, data parallelism (\textit{\sysname{} DP}) and pipeline parallelism (\textit{\sysname{} PP} with \textit{\wap}), while offering users the illusion of running on a single virtual device with practically unbounded memory.
With a user-specified parallelism mode, \sysname{}'s Scheduler binds tasks to devices, appropriately scheduling the movement of required inputs (activations, weights, etc.) from CPU to GPU memory or directly between GPU memories.

\begin{figure}[!t]
  \centering
  \includegraphics[width=\linewidth]{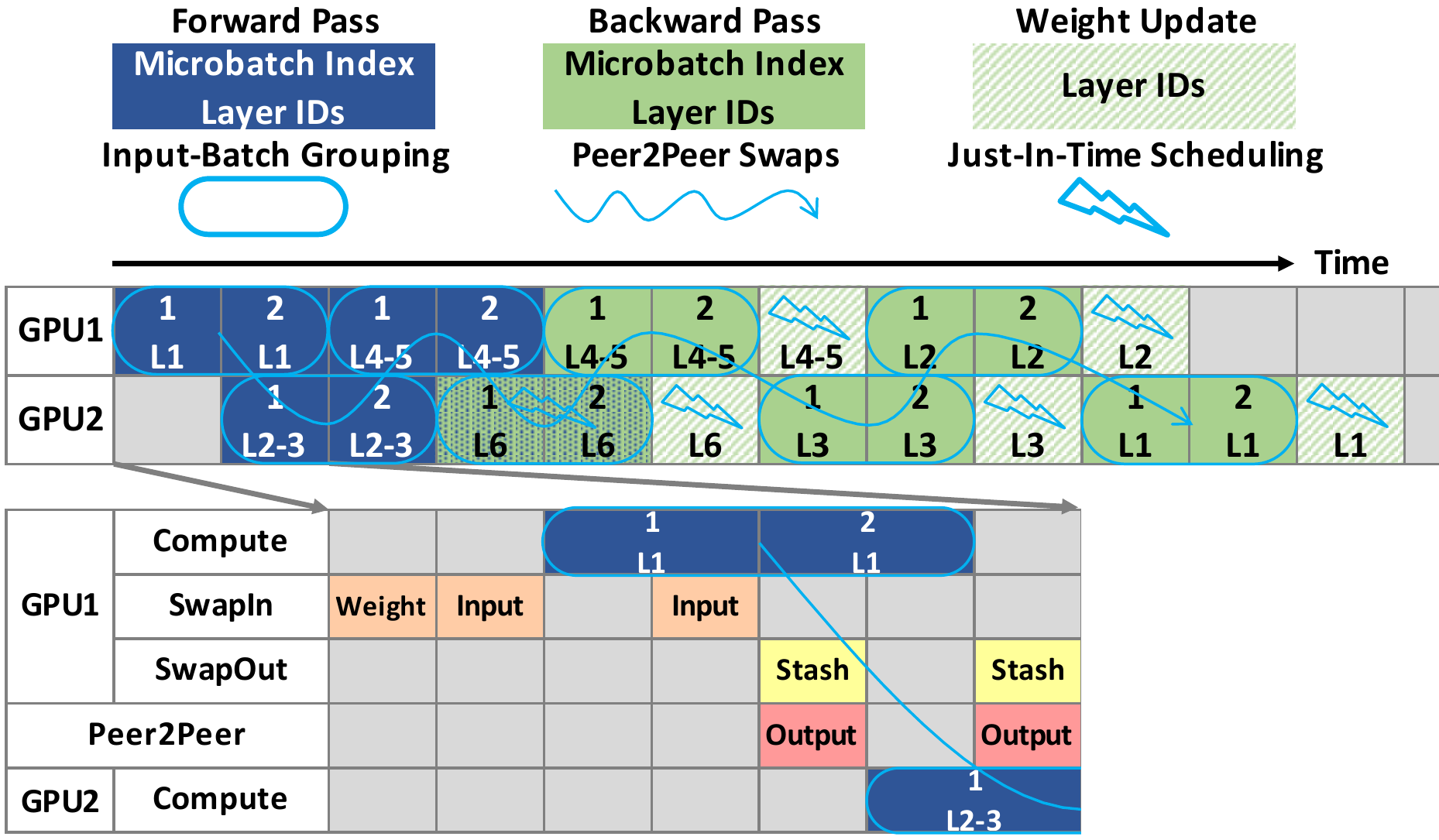}
  \vspace{-4ex}
  \caption{Example of training a toy six-layer ``massive'' model on 2 GPUs with  \wap in \sysname{}.}
  \label{fig:vPP}
  \vspace{-2ex}
\end{figure}
\begin{figure*}[!t]
\centering
\includegraphics[width=\linewidth]{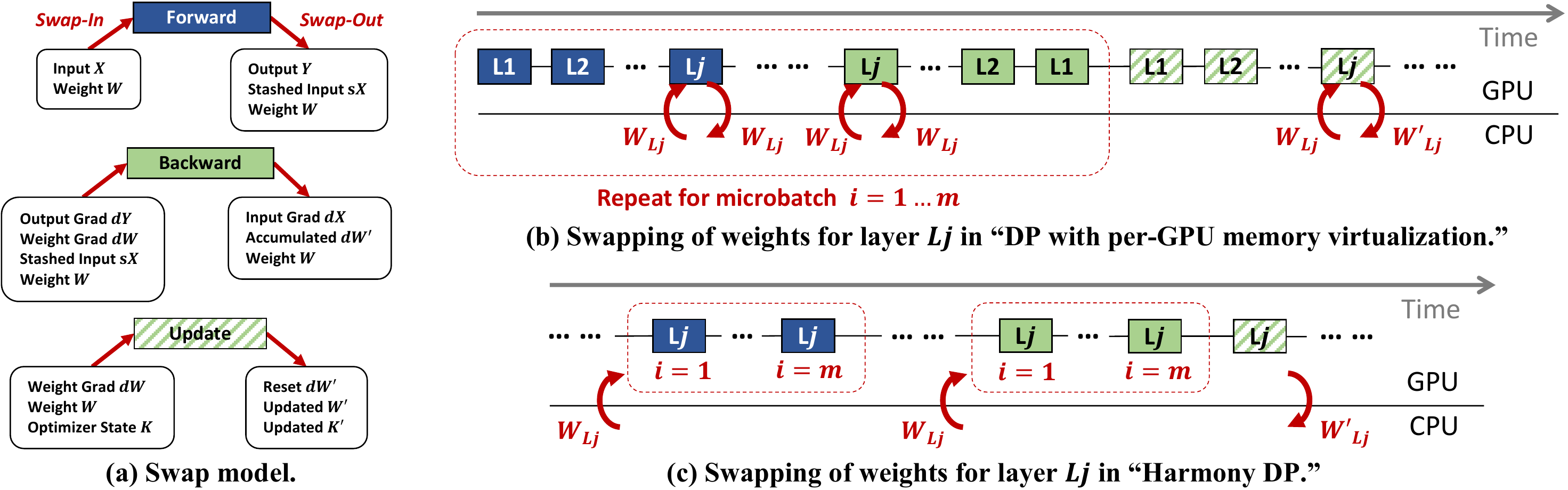}
\vspace{-5ex}
\caption{Tensors that need to be swapped in and out for forward, backward, and weight update phases of training.}
\label{fig:analysis}
\end{figure*}
\niparagraph{Key Optimizations.} Operating at peak throughput requires \sysname{} to overcome two main challenges: (i) minimizing expensive CPU-GPU memory swaps, and (ii) balancing load across all GPUs so that there is no bottleneck worker in the execution pipeline.
\sysname{} achieves this by using four distinct optimizations:

\vspace{0.5ex}\noindent\circled{1} \emp{Input-batch grouping} allows a scheduled layer pack to execute across different input batches back-to-back;  the state of layer(s) (e.g., the weight or gradient buffer) can stay in memory and be reused across multiple input data batches or input tensors.
Grouping $M$ inputs for a layer pack (each input-batch saturates GPU memory) reduces what would otherwise have been $M$ repeated swaps of the state for each batch to a single swap.
Figure~\ref{fig:vPP} shows an example of training with \sysname{} PP, where each layer pack executes on a group of two microbatches back-to-back before moving to the next layer pack.
Unlike traditional pipeline stages~\cite{narayanan2019pipedream, huang2018gpipe} which execute all layers in the stage one batch at a time, resulting in repeated swaps when used with GPU memory virtualization, in \sysname{} the forward pass of a layer (e.g., $L1$) runs through 2 input batches without swapping out its weights, and its backward pass computes the gradient of 2 batches without swapping out its gradient buffer.

\vspace{0.5ex}\noindent\circled{2} \emp{Just-in-time scheduling} executes a task as soon as all its input tensors are available in GPU memory, avoiding delays in execution that risk unnecessarily swapping out the required input tensors, and then swapping them back in later.
For example, in Figure~\ref{fig:vPP} \textit{jit-scheduling} brings the update task of each layer closer to its backward pass so that the weight and gradient tensors needed by the update tasks can be reused while they are still resident in GPU memory (\textit{jit-update}). 
Similarly, the backward pass for the last layer ($L6$) can be scheduled immediately along with its forward pass for each microbatch (\textit{jit-compute}), an optimization especially useful when it avoids the overheads of recomputation for the last layer.

\vspace{0.5ex}\noindent\circled{3} \emp{Generalized p2p swaps} replaces CPU-GPU swaps, for \textit{all} tensors (rather than only the per-stage output in prior work~\cite{narayanan2019pipedream, narayanan2020memory}) that are shared across two layers, with fast device-to-device swaps where applicable.
For the example in Figure~\ref{fig:vPP}, all input and output tensors of layers are transferred directly between the two GPUs.

\vspace{0.5ex}\noindent\circled{4} \emp{Multi-dimensional layer packing} packs together multiple layers executing on a microbatch of input (e.g., forward pass, backward pass, or weight update).
Consequently, both the pack size and the microbatch size of a task determine its memory footprint and performance.
Prior work 
fixes one or both of these parameters, invariably punting the problem to model developers~\cite{shoeybi2019megatron,huang2018gpipe,narayanan2019pipedream}.
\sysname{}'s Configuration Search Engine searches through separate layer packs for the forward and backward pass and their corresponding microbatch sizes to find the best training time configuration that balances compute, memory, and swaps.

\niparagraph{\wap.}
These techniques taken together result in a completely novel pipeline schedule in \sysname{} PP compared to prior work~\cite{narayanan2019pipedream,narayanan2020memory,huang2018gpipe}.
Like GPipe and PipeDream-Flush~\cite{narayanan2019pipedream}, \sysname{} PP also flushes the pipeline at the iteration end, thus providing synchronous SGD semantics.
Unlike prior work that pins layers to GPUs (and with each GPU executing only one layer pack in both the forward and backward pass), each GPU in \sysname{} PP ends up executing \textit{different forward and backward layer packs} enforced by the deterministic wrap-around schedule (e.g., in Figure~\ref{fig:vPP}, GPU1 ends up executing $L1$'s forward and $L2$'s backward pass).
Binding of tasks across $N$ devices in the wrap-around schedule, at a high level, can be described by the following pseudocode:
\begin{mdframed}[backgroundcolor=light-gray,roundcorner=10pt,leftmargin=0, rightmargin=0, innerleftmargin=1, innertopmargin=5,innerbottommargin=5, outerlinewidth=1, linecolor=light-gray]
\begin{lstlisting}[language=Python]
  // Assumption: Task($P_B$[i]) also performs wt. updates
  $P_{FB}$ = $P_F$ + Reverse($P_B$)
  for i in range($P_{FB}$):
      Task($P_{FB}$[i]) $\rightarrow$ GPU[i $\bmod$ $N$] // bind task to GPU
\end{lstlisting}
\end{mdframed} 
Furthermore, with per-GPU memory virtualization, prior approaches have to repeatedly swap out and then swap back in weights and gradients of layers while executing across microbatches (data parallelism and PipeDream's 1F1B); by contrast, \sysname{} PP \textit{groups} the executions of a layer pack \textit{across all microbatches} in a minibatch before scheduling the next layer pack on that GPU.

\niparagraph{Intuitive Example to Highlight Advantages.}
To explain how \system{} significantly reduces swap overhead, using a simplified example we provide an analytical comparison between \sysname{} and the corresponding baselines that use per-GPU memory virtualization.
We assume (without loss of generality) a setup with homogeneous GPUs where each GPU's memory capacity permits it to only hold one layer operating on one microbatch at any time.  
We also assume a simplified DNN model with one type of layer (like Transformers) and where each layer has the same runtime and memory footprint for its forward, backward, and update phases.

\sysname{} provides generalized support for swapping all tensors across different layers where they each need to swap in/out certain inputs/outputs (Figure~\ref{fig:analysis}(a)).
First, we focus on a specific kind of tensor, model weights $W$ (with a size of $|W|$), to provide an intuition for such reductions in swap overhead when training a model of $R$ layers (i.e., $|W| = \sum_{j=1}^{R} |W_{Lj}|$) with $m$ microbatches per GPU and $N$ GPUs (for a minibatch of $mN$ microbatches).
Figure~\ref{fig:analysis}(b) shows that, for a single iteration (minibatch), when using DP with per-GPU memory virtualization, \textit{each GPU} has to swap $W$ in and out for both the forward and backward passes independently and this has to be done for \textit{each of the $m$ microbatches}. 
At the end of the iteration, each GPU also has to swap $W$ in and out once for weight update.  
This results in an overall swap volume of $(4m + 2)N|W|$ per iteration. 
By contrast, in \system{} DP (Figure~\ref{fig:analysis}(c)), each GPU has to swap $W$ in \textit{only once} each for the forward and the backward passes \textit{across all $m$ microbatches} (due to \textit{input-batch grouping}), and swap $W$ out once for weight update (due to \textit{jit-scheduling}), resulting in an overall swap volume of $3N|W|$ per iteration. 

The same swap analysis also applies to PP with per-GPU memory virtualization. 
But the key difference is that PP does not have duplicated weight per GPU, canceling the $N$ term in the swap volume, i.e., $(4m + 2)|W|$.
Finally, \system{} PP (Figure~\ref{fig:vPP}) combines the best of the two worlds with both \textit{input-batch grouping} and no duplicated weights, bringing the overall per-iteration swap volume down to $3|W|$ (across all $m$ microbatches and all $N$ GPUs)!

For brevity, here we omit a full analytical comparison for all tensors shown in Figure~\ref{fig:analysis} and refer the reader to the Appendix for more details; suffice to say, \sysname{} offers swap reduction for all tensors and \sysname{} PP dominates reductions in swap volume. We empirically show the advantages of \sysname{} in \secref{sec:eval}.

\section{Design and Implementation}
\label{sec:impl}
\sysname{} is implemented in Python (54K LOC) on top of PyTorch.  
Next, we present the details of \sysname{}'s components in Figure~\ref{fig:overview}.

\subsection{Decomposer}
\label{sec:decomposer}

\begin{figure}[!t]
  \centering
  \includegraphics[width=0.7\linewidth]{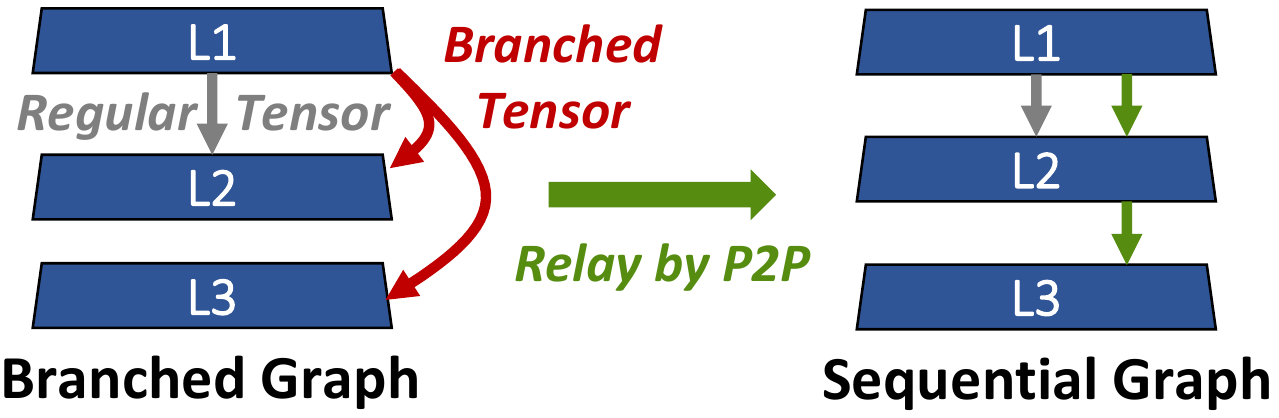}
  \vspace{-1em}
  \caption{Serializing layer graphs in \sysname.}
  \label{fig:sequential}
  \vspace{-2ex}
\end{figure}

\sysname{}'s Decomposer constructs a fine-grained layer graph from an imperative-style PyTorch script and generates code so that each layer can be executed individually.
The main challenge is dealing with branching in the model.  
\sysname{} overcomes this issue by relaying the branch tensor across downstream layers using p2p swaps until the destination layer consumes, thus minimizing CPU-GPU swaps.
We have implemented such a \textit{p2p-relaying} scheme to serialize the layer-level graph by adding identity nodes across layers as shown in Figure~\ref{fig:sequential}.


Unlike prior approaches that generate code for entire pipeline stages and bind them to a GPU early (e.g., PipeDream~\cite{narayanan2019pipedream}), \sysname{} Decomposer uses the layer graph to generate code such that each layer can be invoked individually, and it delays layer packing and GPU binding to the downstream \sysname{} Scheduler.

\subsection{Profiler}
\label{sec:profiler}
With the generated layer code, using a single GPU, \sysname{}'s Profiler runs each layer individually and records profiles: compute time, memory footprint, and input tensor size.
Since \sysname{} tunes both microbatch size and layer packs for both forward and backward pass, we also need to collect profiles for each layer under different microbatch sizes.
Brute-force profiling with every possible microbatch size is impractical.
Instead, \system{} 
sweeps through microbatch sizes to determine the maximum microbatch size that does not cause out-of-memory problems by using a process similar to TCP slow start (multiplicative increase of microbatch size, halving at the first OoM, and then additive increase until the next OoM). 
It then profiles layers for each microbatch size from 1 to this max size at fixed stride intervals. Finally, \sysname{} uses a simple regression model to interpolate each layer's characteristics for microbatch sizes that it does not sample. 
We validate the efficacy of the final profiling estimation, 
showing that it is strikingly accurate.

\subsection{Scheduler}
\label{sec:scheduler}

Using the layer-granularity profiles and machine model, \sysname{}'s Scheduler searches through the space of training configurations, estimating iteration time for each configuration and picking the fastest among them for execution by the Runtime.

\begin{figure}[!t]
  \centering
  \includegraphics[width=\linewidth]{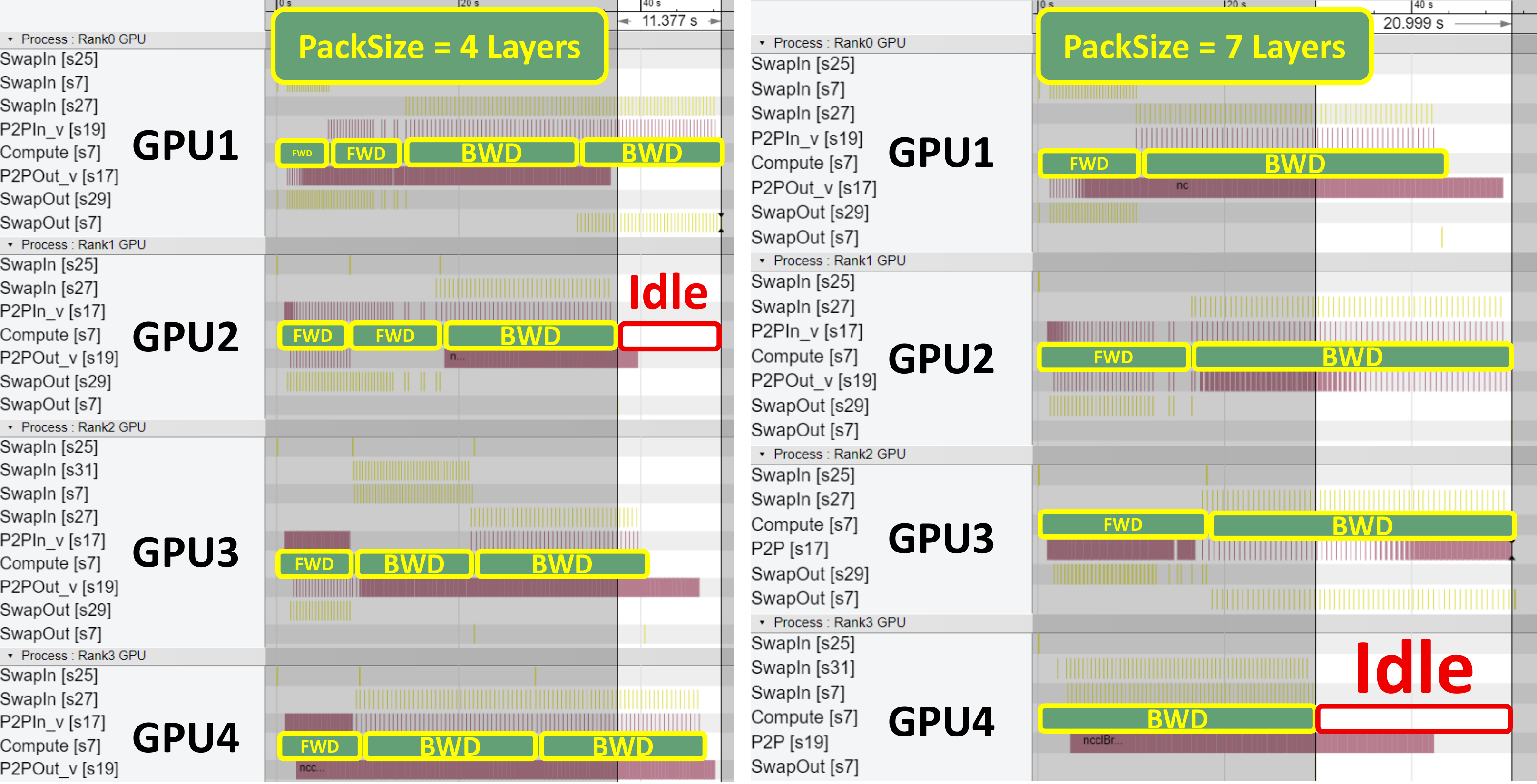}
  \vspace{-4ex}
  \caption{Greedily packing more layers to satisfy only memory capacity constraints can cause greater load imbalance across GPUs due to coarser-granularity tasks. Configurations: left: every 4 layers form a pack, $U_F = 30, U_B = 15$; right: every 7 layers form a pack, $U_F = 20, U_B = 10$.}
  \label{fig:pack-size-diff}
  \vspace{-2ex}
\end{figure}
\subsubsection{Configuration Search Engine}
\label{sec:config_search}
\hfill\\\noindent
We define a configuration to be a four-tuple: \textbf{\textit{$<$forward microbatch size $U_F$, forward layer packs $P_F$, backward microbatch size $U_B$, backward layer packs $P_B {>}$}}.
Unlike prior work~\cite{shoeybi2019megatron,huang2018gpipe,narayanan2020memory,li2014scaling}, which either assumes 
the microbatch size to be specified by the user, or fixes the microbatch size and layer packs to be the same between the forward and the backward pass, \sysname{}'s Configuration Search automatically determines the entire four-tuple.
We expect users to specify a mini-batch size (not the microbatch size) as it directly affects convergence~\cite{krizhevsky2014one,goyal2017accurate,hoffer2017train,you2017lars}.
But determining the
four-tuple above is challenging for a number of reasons.

\begin{algorithm}[!t]
\DontPrintSemicolon
\newcommand\mycommfont[1]{\footnotesize\ttfamily\textcolor{blue!50!black}{#1}}
\SetCommentSty{mycommfont}
\SetKwInOut{Input}{Input}
\SetKwInOut{Output}{Output}
\caption{\sysname Configuration Search}
\label{alg:config_search}
\Input{number of layers $R$, minibatch size $D$,\\
       maximal forward microbatch size $U_{FMAX}$,\\
       maximal backward microbatch size $U_{BMAX}$,\\
       adopted packing method $\lambda$ (returns layer packs $P$),\\
       profiled time/memory/activation size $\phi$,\\
       GPU memory capacity $\alpha$, PCIe bandwidth $\beta$,\\
       \sysname mode $H$, number of GPUs $N$,\\
       task graph generator $\rho$, runtime estimator $\varepsilon$}
\Output{best configuration $(U_F^*,\ P_F^*,\ U_B^*,\ P_B^*)$}
\tcp{find effective maximal microbatch size}
\If{$H$ is ``Harmony DP''}
{ $D \gets D/N$\; }
$U_{FMAX},\ U_{BMAX} \gets \min(U_{FMAX},\ D),\ \min(U_{BMAX},\ D)$\;
\tcp{search for best config. with minimal time}
$(U_F^*,\ P_F^*,\ U_B^*,\ P_B^*) \gets None$ \tcp*[l]{best configuration}
$t^* \gets \infty$ \tcp*[l]{best runtime}
\For{$U_B \gets 1$ \textbf{to} $U_{BMAX}$} 
{
    $P_B \gets \lambda(\text{``}B\text{''},\ U_B,\ R,\ \phi,\ \alpha)$ \tcp*[l]{backward packing}
    \For{$U_F \gets 1$ \textbf{to} $U_{FMAX}$}
    {
        $P_F \gets \lambda(\text{``}F\text{''},\ U_F,\ P_B,\ \phi,\ \alpha)$ \tcp*[l]{forward packing}
        \tcp{from current config., generate task graph}
        $G \gets \rho(U_F,\ P_F,\ U_B,\ P_B, \ H,\ N,\ D)$\; 
        $t \gets \varepsilon(G,\ H,\ N,\ \phi, \ \beta)$ \tcp*[l]{estimate runtime}
        \If{$t < t^*$}
        {
            $(U_F^*,\ P_F^*,\ U_B^*,\ P_B^*) \gets (U_F,\ P_F,\ U_B,\ P_B)$\;
            $t^* \gets t$\;
        }
    }
}
\KwRet{$(U_F^*,\ P_F^*,\ U_B^*,\ P_B^*)$}
\end{algorithm}
\begin{algorithm}[!t]
\DontPrintSemicolon
\newcommand\mycommfont[1]{\footnotesize\ttfamily\textcolor{blue!50!black}{#1}}
\SetCommentSty{mycommfont}
\SetKwInOut{Input}{Input}
\SetKwInOut{Output}{Output}
\caption{Balanced Time Packing $\lambda$}
\label{alg:bt_pack}
\Input{forward or backward type $\tau$, microbatch size $U$,\\
       number of layers to pack $R$ (or given packs $P_B$),\\
       profiled time/memory/activation size $\phi$,\\
       GPU memory capacity $\alpha$}
\Output{layer packs $P$}
\If{$P_B$ exists}
{ $R \gets P_B.RemoveLastPack().CountLayers()$\tcp*[l]{jit compute} }
$t \gets \phi(\tau, U, R).PerLayerTimeList()$\;
$m \gets \phi(\tau, U, R).PerLayerMemoryList()$\;
\tcp{loop num of packs from the smallest (largest packs)}
$S_{min} \gets m.Sum()/\alpha$\;
\For{$S \gets S_{min}$ \textbf{to} $R$}
{
    \tcp{find packs with per-pack time closely equal}
    $c \gets t.Sum()/S$\tcp*[l]{average per-pack time}
    $c' \gets [c,2c,\ldots,(S-1)c]$\tcp*[l]{accumulated pack times}
    $t' \gets t.PrefixSum()$\tcp*[l]{accumulated layer times}
    $i \gets BinarySearch(t',\ c')$\tcp*[l]{insert c' into t' and get insertion points} 
    $P \gets t.Split(i).ToLayerID()$\tcp*[l]{packs found} 
    \tcp{check if any pack is over capacity}
    \For{$p \gets P[0]$ \textbf{to} $P[S-1]$}
    {
        \If{$m[p].Sum() > \alpha$}
        {
            break; continue\tcp*[l]{try smaller packs} 
        }
    }
    \KwRet{$P$}\tcp*[l]{balanced time and largest pack size} 
}
\end{algorithm}

First, both the pack size and the microbatch size of of a task determine the memory footprint and performance when executing the task.
It is not immediately clear if one should maximize the microbatch size or the layer pack size to maximally utilize the memory capacity  of a device.
Given a fixed memory capacity, increasing the pack size can reduce p2p and CPU-GPU swap volume (especially when using recompute~\cite{chen2016training}).
Unfortunately, greedily constructing as large a pack as can fit the memory of individual GPU results in globally sub-optimal pipelines.
Figure~\ref{fig:pack-size-diff} shows such an example of training a BERT-Large with \system{} PP; the configuration with larger packs and smaller microbatch size (right) results in load imbalance across GPUs and up to $2\times$ longer idle times than a configuration with smaller packs and larger microbatches (left).

Second, while it might be tempting to identify only backward packs and microbatch sizes, and reuse
them for the forward pass (a scheme we term \textit{Equi-FB}), this is far from optimal because the forward and the backward pass for the same layer can have very different characteristics.  For example, it is common for the backward pass for a layer to have $2\--3\times$ the runtime and memory footprint of the forward pass, thus motivating the need for different pack and microbatch sizes across these passes.  Our experiments show that \textit{Equi-FB} is 30\% slower than picking separate values for forward and backward packs and microbatch sizes in a four-tuple configuration.

\niparagraph{Heuristic-based Search.}
The problem of finding the optimal configuration that minimizes the training time can be shown to be NP-hard\footnote{We omit the hardness proof here for brevity, but refer the interested reader to the Appendix for more details.}, which makes it unlikely that we can find a provably optimal configuration efficiently. 
We address this challenge by using a simple but effective heuristics-based search algorithm (Algorithm~\ref{alg:config_search}) to identify a high-performance four-tuple configuration.  We proceed roughly as follows:
\begin{itemize}[leftmargin=*,itemsep=0mm]

\item 
We first determine the backward layer packs $P_B$ for each backward microbatch size $U_B$ (Lines 6, 7). This helps us identify the input tensors of each pack in $P_B$ that we need to checkpoint in the forward pass; these input tensors will be used to recompute stashed tensors for all intermediate layers in the pack before we start the backward-pass compute for the task~\cite{chen2016training}.  We can then use this information in determining the forward layer packs $P_F$ for each forward microbatch size $U_F$ we sweep through (Lines 8, 9). 
Furthermore, the last layer pack is shared between $P_F$ and $P_B$, 
avoiding recompute for the first backward task (jit-compute, Line 2 of Algorithm~\ref{alg:bt_pack}).

\item To reduce load imbalance across GPUs and avoid stragglers in the pipeline,
we propose a method to determine layer packs that balances the time taken by each pack while maximizing average pack size.
Algorithm~\ref{alg:bt_pack} outlines our method, which runs in time $\mathcal{O}(R^2)$ (invoked by Algorithm~\ref{alg:config_search} at Lines 7, 9).

\item For each configuration ($U_F$, $P_F$, $U_B$, $P_B$) to be explored, we generate a task graph, \textit{binding} each task to an individual GPU (Algorithm~\ref{alg:config_search}, Line 10). 

\item We then \textit{estimate} the end-to-end runtime of an iteration for a task graph (Algorithm~\ref{alg:config_search}, Line 11).  
The estimation leverages profiles of individual layers ($\phi$) from Profiler, and performs an \textit{event-driven simulation} to capture swap, transfer, and compute times.  Simulating an iteration without actually running it on real hardware enables fast configuration search.  Later, in evaluation, we show that these estimated times closely match real end-to-end runs (see Figure~\ref{fig:est_acc}).

\item The search returns the configuration with the best iteration time in the set of configurations explored (Algorithm~\ref{alg:config_search}, Lines 12--15).

\end{itemize}
In total, the time complexity of all steps in Scheduler (heuristic-based search, balanced time packing, task graph generation, runtime estimation) is $\mathcal{O}(U_{FMAX} \cdot U_{BMAX} \cdot R (R+D))$, where $U_{FMAX/BMAX}$ is the maximal possible forward/backward microbatch size, $R$ is the number of layers, and $D$ is the minibatch size. 
In practice, this end-to-end scheduling time is less than 32 seconds (see Table~\ref{tab:search_time}).

\subsubsection{Task Graph Generation}
\label{sec:tgg}
\hfill\\\noindent
A \textbf{task} is the unit of execution in \sysname. 
Figure~\ref{fig:vPP} shows the three types of tasks: \textit{Forward}, \textit{Backward}, and \textit{Weight Update}.
Each task is associated with a layer pack and a microbatch size (the result of configuration search).
Each task is bound to a specific execution backend (GPU device or CPU process).
For instance, the second task in Figure~\ref{fig:vPP}, bound to GPU\#2, is a \textit{Forward} task for layer pack \textit{[2,3]} with $U_F$=10 and a group of two microbatches.
Each task also specifies the required inputs and outputs to be swapped in and out, respectively, along with the channels they ought to be transmitted on.
For the same example, the task specifies its two inputs: tensor \textit{L1 Output} over a \textit{Peer2Peer} input channel, and \textit{L2-3 Weight} tensors over a \textit{CPU-GPU Swap} channel.
The complete list of inputs/outputs to be swapped is shown in Figure~\ref{fig:analysis}(a), where each input/output can choose from one of the four channels: CPU-GPU Swap, Peer2Peer, Message Passing, and Shared Memory.
Putting together the tasks of an entire iteration results in a \textbf{task graph}, where each node is a task and each edge is the specified input/output between tasks.
Such task graphs are used to drive the \sysname{} Runtime. 

\niparagraph{\sysname DP and PP.} 
Using the task graph, \sysname{} is able to schedule a variety of distributed training schedules; it does this by unrolling an iteration's tasks across GPUs.  
\sysname{} can support conventional Data Parallel and Pipeline Parallel training, both enhanced with per-GPU memory virtualization (that we hereafter call per-GPU swap).
Crucially, it supports two new schedules, \sysname{} DP (Figure~\ref{fig:analysis}) and \sysname{} PP (Figure~\ref{fig:vPP}) \--- both these schemes benefit from \sysname{}'s four key optimizations. 

\begin{figure}[!t]
  \centering
  \includegraphics[width=\linewidth]{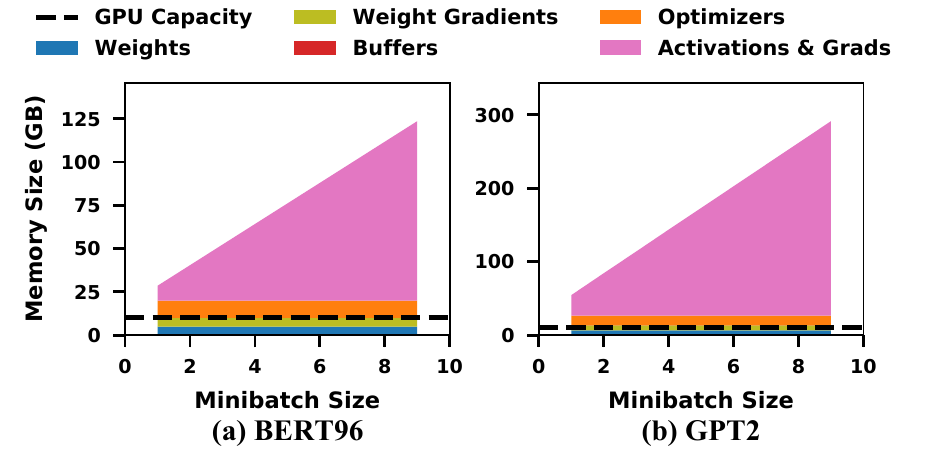}
  \vspace{-6ex}
  \caption{Memory footprint statistics for training massive models (using virtualized GPU memory).}
  \label{fig:memory_scale}
  \vspace{-2ex}
\end{figure}
\subsection{Runtime}
\sysname{}'s Runtime executes in CPU processes, one for each GPU in the deployment.
This 1:1 mapping is required to enable effective concurrency and overcome the limitations of the Python GIL.  
Each runtime process can also be pinned to a CPU core on a socket which has NUMA affinity to the GPU it controls.
All tasks run on GPUs, but \sysname{} also supports the \textit{Weight Update} task being offloaded to the CPU.
Each runtime process executes the ordered list of tasks in the unrolled task graph handed to the Runtime by the Scheduler, and repeats it for all training iterations.

\niparagraph{CUDA Streams and CUDA Events.}
To effectively utilize the GPU and overlap computation and communication, \sysname{} uses \textit{5 distinct CUDA streams}: one each for compute, swap-in, swap-out, p2p-in, and p2p-out on every GPU.  We use CUDA events across streams to synchronize for task dependencies.
The swap and p2p streams are managed by background CPU runtime threads for pre-allocating CPU tensors, prefetching GPU tensors for upcoming tasks, waiting for swap completion and tensor transfers.
Prefetching uses extra GPU memory to overlap swaps/transfers with compute and uses double buffering to avoid repeated allocations.

\niparagraph{Memory Manager.}
In PyTorch, each CUDA stream can allocate, free, and reuse its own memory. While streams can share memory buffers, memory reuse is private to each stream (e.g., the memory freed by stream A is not reusable by stream B); such private reuse can shrink the effective memory available to individual streams.
To overcome this limitation, \sysname{}'s Runtime employs a \textit{``central'' memory manager} on the compute stream and allows it to manage memory for all streams with an unified memory pool.
%

\section{Evaluation}
\label{sec:eval}
\subsection{Experimental Setup}
\label{sec:eval_setup}
\niparagraph{Configurations.}
We run experiments on three server configurations. 
Two of them are commodity servers with four and eight GTX-1080Ti GPUs (11 GB each)~\cite{ASUS, GTX}, and 18-core (375GB DRAM) and 36-core (750GB DRAM) 2.3GHz Xeon CPUs~\cite{Xeon}, respectively.  
The third server is an NVIDIA DGX-2 with 16 V100 GPUs (32GB each), 96-core Xeon CPU (1.5TB DRAM), and NVSwitch~\cite{DGX2,NVLink}.
On all servers, GPUs are connected to CPU via a PCIe tree as in Figure~\ref{fig:pcie}, where each link is a 16-lane PCIe3 (16GB/s per direction).
All results shown are with PyTorch 1.5, NCCL 2.4, CUDA 10.1, and FP32 precision.
\textit{Unless explicitly stated, we show evaluation results on the commodity server configuration with four GTX-1080Ti GPUs}.
\begin{figure*}[!t]
  \centering
  \includegraphics[width=0.95\textwidth]{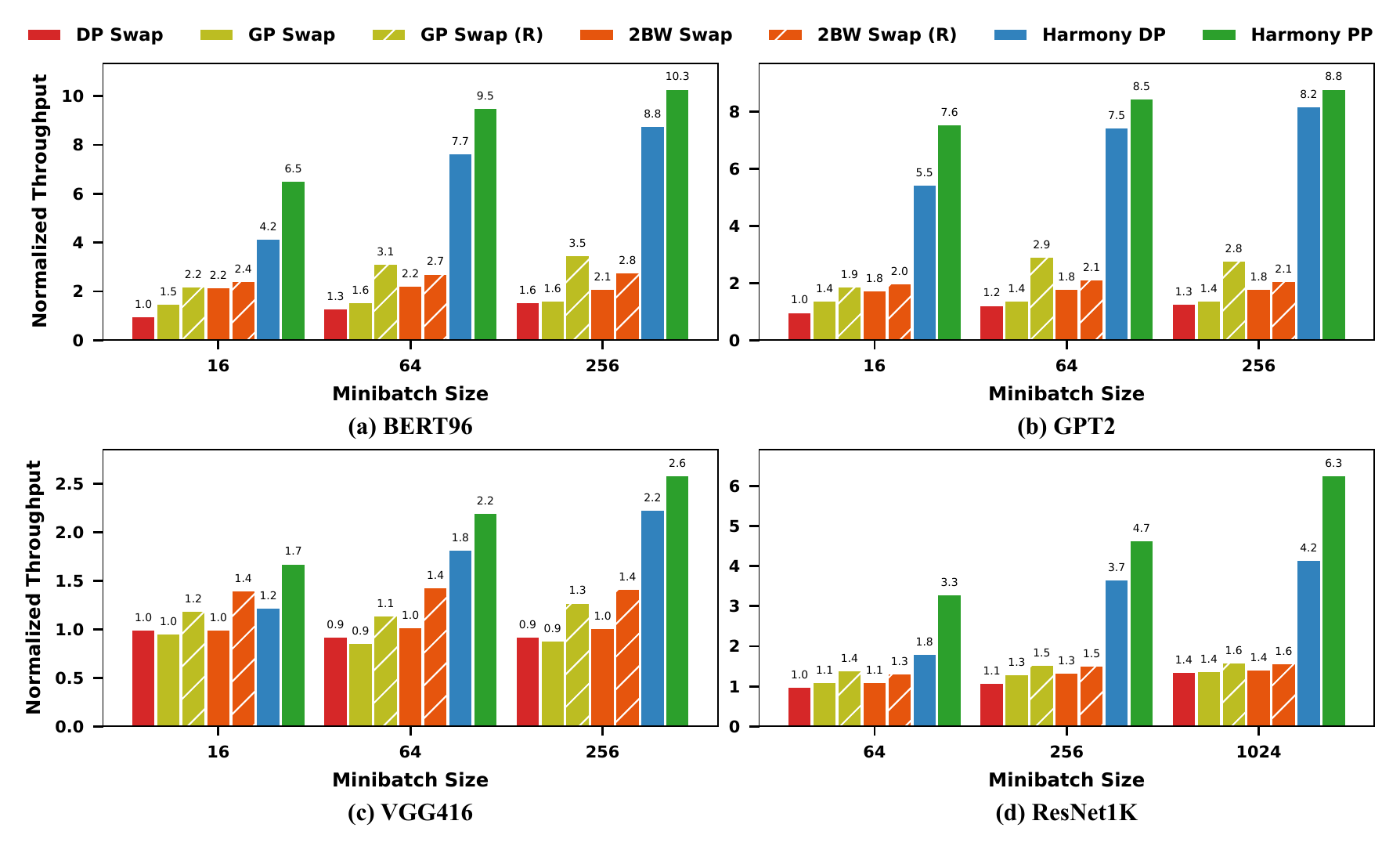}
  \vspace{-4ex}
  \caption{Performance comparison with per-GPU swap baselines by training different models with various minibatch sizes on 4 GPUs. Each group of bars represents one minibatch size. \textit{R} denotes the usage of recompute for activations. Throughput is normalized against \dps at the smallest minibatch size (i.e., the leftmost bar).}
  \label{fig:throughput_vs_baselines}
\end{figure*}

\niparagraph{Models.}
Our evaluation uses the following DNN models:
\begin{itemize}[leftmargin=*,itemsep=0mm]
    \item Two BERT variants: \textbf{BERT-Large} (24 transformer layers)~\cite{devlin2018bert} and \textbf{BERT96} (96 transformer layers)~\cite{narayanan2020memory}. 
    Both models use a sequence length of 512 and training uses the GLUE dataset~\cite{wang2018glue} with an Adam optimizer.
    \item Three GPT2 variants: \textbf{GPT2} (the default model with 1.5B parameters)~\cite{radford2019gpt2}, \textbf{GPT2-Medium} (0.3B)~\cite{BlogGPT2}, and customized GPT2 models (\textbf{10s Billion})~\cite{rajbhandari2021zeroinf}.
    All are trained with a sequence length of 1024 on WikiText dataset~\cite{merity2016pointer} and an Adam optimizer.
    \item \textbf{VGG416}. This is a variant of the classic VGG model scaled to have 416 layers and has been used for evaluating per-GPU memory virtualization in prior work~\cite{krizhevsky2012imagenet,rhu2016vdnn,hildebrand2020autotm}.  It is benchmarked for training using the ImageNet~\cite{li2009imagenet} dataset with a SGD optimizer.
    \item \textbf{ResNet1K}. Another CNN, a ResNet variant~\cite{he2016resnet,he2016resnetv2}, used for evaluating per-GPU memory virtualization in prior work~\cite{chen2016training,jin2018layrub}.
\end{itemize}

\niparagraph{Memory Footprint.}
The working set size of these models exceeds the combined memory capacity of our GPUs; the memory footprint far exceeds the capacity of an individual GPU for even the smallest batch sizes.
Figure~\ref{fig:memory_scale} analyzes the memory footprint of training two massive models at different batch sizes and also breaks down the memory footprint into important components (weights, gradients, etc.). 
The memory footprint analysis of other models is similar.

\niparagraph{Per-GPU Swap Baselines.}
Given the prohibitive memory requirements mentioned above,
we enhance existing approaches for parallel DNN training, such as Data Parallelism (DP)~\cite{li2020ddp}, GPipe (GP)~\cite{huang2018gpipe}, and PipeDream-2BW (2BW)~\cite{narayanan2020memory}, to incorporate per-GPU memory virtualization using IBM-LMS~\cite{PTLMS,TFLMS}.  
As a result, we construct new viable baselines for comparison to \sysname{}: \textbf{\dps}, \textbf{\gps}, and \textbf{\pds}.
Furthermore, we augment these baselines with memory optimizations: 1) \textit{gradient accumulation}~\cite{narayanan2020memory}, 2) \textit{inplace operations} and \textit{memory reuse}~\cite{chen2016training,wang2018superneurons}, and 3) other memory buffer optimizations. 
While \sysname{} always uses recompute~\cite{chen2016training} to cut the memory of stashed activations, we also enable \gps and \pds to use recompute, thus creating additional baselines \textbf{\gpsr} and \textbf{\pdsr} respectively.

\niparagraph{ZeRO-Infinity.}
We also compare against ZeRO-Infinity~\cite{rajbhandari2021zeroinf}, a recent swap-based DP enhancement that supports moving state between CPU and GPU memory, offloads weight update to CPU, and shards model state across GPUs only to swap in every layer's state when required for (re)compute on each GPU. 
ZeRO-Infinity also includes NVMe storage devices in the memory hierarchy, if available (e.g., high-end DGX-2 servers~\cite{DGX2}).
In this paper, we only consider massive models whose working set fits in CPU memory and thus \sysname{} does not use storage devices for swaps (many commodity servers lack fast NVMe devices).

\niparagraph{Goal.} We seek to answer the following questions in evaluation: 
\vspace{-1.0ex}
\begin{itemize}[leftmargin=*,itemsep=0mm]
    \item How does \sysname{} compare to baselines, with respect to training throughput and swap overhead? (\secref{sec:compare_baselines}-\ref{sec:compare_zeroinf})
    \item Is \sysname{} training correct (converges as baseline)? (\secref{sec:correctness})
    \item How much does each of our optimizations contribute? (\secref{sec:breakdown_of_optimizations})
    \item How does \sysname{}'s Scheduler perform? (\secref{sec:scheduler_effectiveness})
    \item How does \sysname{} scale with model sizes and GPUs? (\secref{sec:scalability_1080ti})
\end{itemize}

\begin{figure*}[!t]
  \centering
  \includegraphics[width=0.95\linewidth]{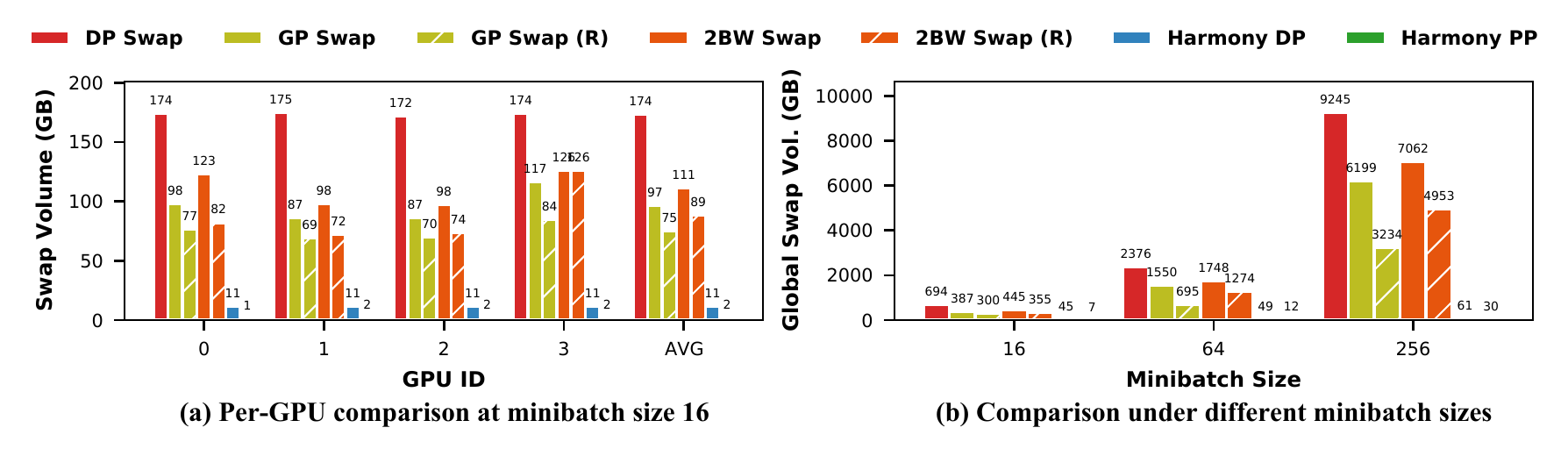}
  \vspace{-4ex}
  \caption{CPU-GPU swap volume comparison of different approaches for training GPT2 on 4 GPUs. Swap volume is measured per minibatch. Global swap volume aggregates swap volume across all GPUs.}
  \label{fig:swapvol_gpt2}
  \vspace{-3ex}
\end{figure*}
\begin{figure*}[!t]
  \centering
  \includegraphics[width=0.95\linewidth]{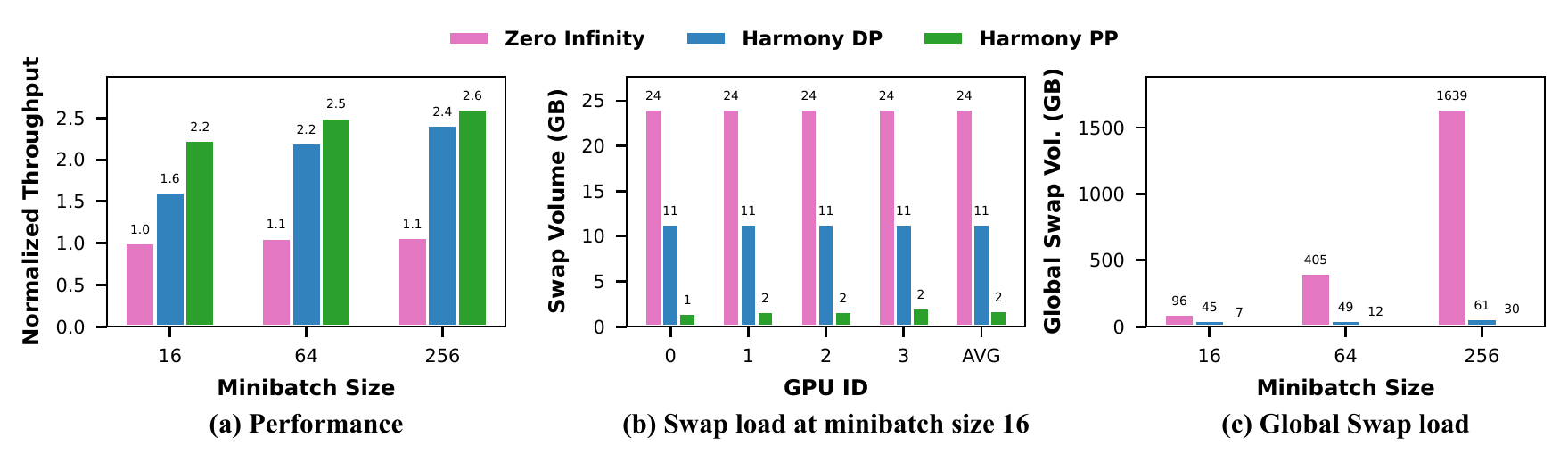}
  \vspace{-4ex}
  \caption{Comparison with \zinf for training GPT2 (1.5B) on 4 GPUs. Throughput is normalized against \zinf at minibatch size 16 (i.e., the leftmost bar). CPU-GPU swap volume is measured per minibatch.} 
  \label{fig:gpt2_vs_zeroinf}
\end{figure*}

\subsection{Comparison with Per-GPU Swap Baselines}
\label{sec:compare_baselines}
Figure~\ref{fig:throughput_vs_baselines} compares \sysname{} with per-GPU swap baselines for different minibatch sizes.
We highlight \emp{five key takeaways}:

First, for any given minibatch size, \emp{\dps consistently underperforms other approaches} -- unsurprisingly, given that each of the 4 GPUs is swapping the entire model state back and forth to CPU memory including unnecessary and repeated swaps across microbatches (\secref{sec:motivation}).  
Figure~\ref{fig:swapvol_gpt2} further reveals that \dps dominates the swap volume over other approaches.

Second, \emp{\gps is consistently worse than \pds} not just due to swap load but also due to pipeline flushes in \gp.
But because swap overheads dominate, the gap between \gps and \pds is less dramatic than when the model fits the collective memory capacity of all GPUs~\cite{narayanan2020memory}.
The baselines using \emp{recompute, \gpsr and \pdsr, perform much better than their no-recompute counterparts (\gps and \pds)} across all models and batch sizes, and this can be directly attributed to the reduced swap overheads due to recompute, which indicates that swap overhead dominates over compute cost.
Figure~\ref{fig:swapvol_gpt2}(a) shows this reduction in swap overheads.

Third, \emp{\sysname DP benefits from input-batch grouping, jit-scheduling, and layer packing, significantly outperforming all baselines} (Figure~\ref{fig:throughput_vs_baselines}), 
with speedups up to $2.4\times\sim7.0\times$ for all models.
\sysname{} DP's swap overheads are an order of magnitude lower than \dps (Figure~\ref{fig:swapvol_gpt2}).

Fourth, \emp{\sysname PP is consistently the fastest approach across all models and minibatch sizes} (Figure~\ref{fig:throughput_vs_baselines}), with speedups up to $2.8\times\sim7.6\times$ over \dps.
It is up to $1.5\times$ faster than \sysname{} DP, further benefiting from pipeline parallelism (eliminating all redundancy in CPU-GPU swaps) and p2p swaps, with a swap volume that is \textit{two orders of magnitude} lower than \dps (Figure~\ref{fig:swapvol_gpt2}).

Fifth, across all models, \emp{\sysname's speedup over baseline approaches widens with larger batch sizes}. This is primarily fuelled by reduced swap load due to \textit{input-batch grouping} in \sysname{}. 
Figure~\ref{fig:swapvol_gpt2}(b) shows that while swap load proportionally goes up for all approaches as we increase batch size, the swap volume across all GPUs is $100\times\sim300\times$ higher for per-GPU swap baselines 
compared to \sysname{}, thus resulting in a flatlining of throughput for baselines (Figure~\ref{fig:throughput_vs_baselines}).

\subsection{Comparison with ZeRO-Infinity}
\label{sec:compare_zeroinf}
We now compare \sysname{} to \zinf on our deployment. 
\zinf suffers from coarse-grained scheduling and lacks optimizations such as input-batch grouping and configuration search for principled layer packing.
For a fair comparison, in our evaluation, we make \zinf share the same configuration as \sysname (i.e., minibatch size, microbatch size, pack size for recompute) that \sysname{} finds and enable all its relevant optimizations.
Figure~\ref{fig:gpt2_vs_zeroinf}(a) shows that \emp{\sysname DP and PP are up to $2.3\times$ and $2.5\times$ faster than \zinf}, respectively, for GPT2. 
\sysname{}'s throughput \textit{speedup widens as the minibatch size increases}.
Figures~\ref{fig:gpt2_vs_zeroinf}(b,c) show that this speedup can be directly attributed to \textit{an order-of-magnitude lower swap load} in \sysname with input-batch grouping (and p2p swaps in \sysname PP).

\begin{figure}[!t]
\centering
    \begin{subfigure}[b]{0.48\textwidth}
      \includegraphics[width=\linewidth]{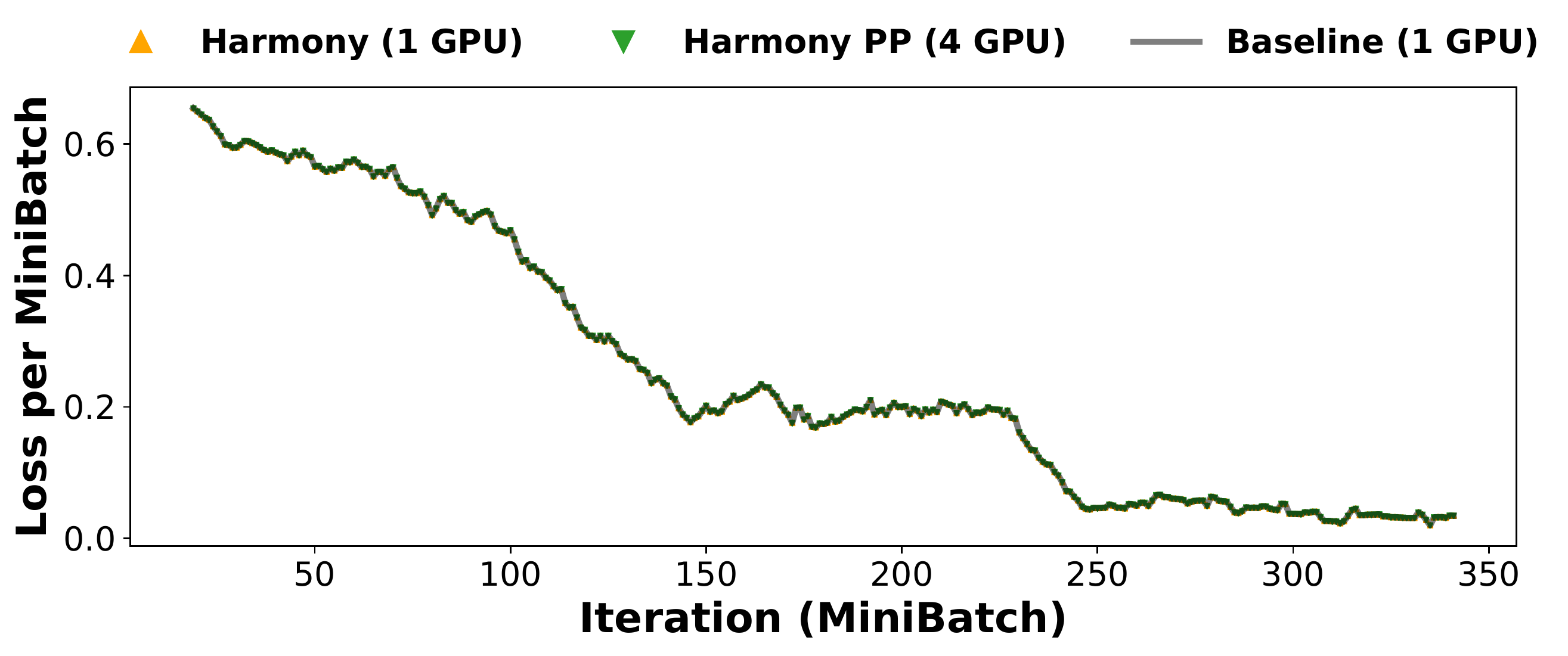}
      \vspace{-4ex}
      \caption{\sysname vs. single-GPU baseline.}
    \end{subfigure}
    \vspace{1ex}
    \begin{subfigure}[b]{0.47\textwidth}
      \centering
      \includegraphics[width=\linewidth]{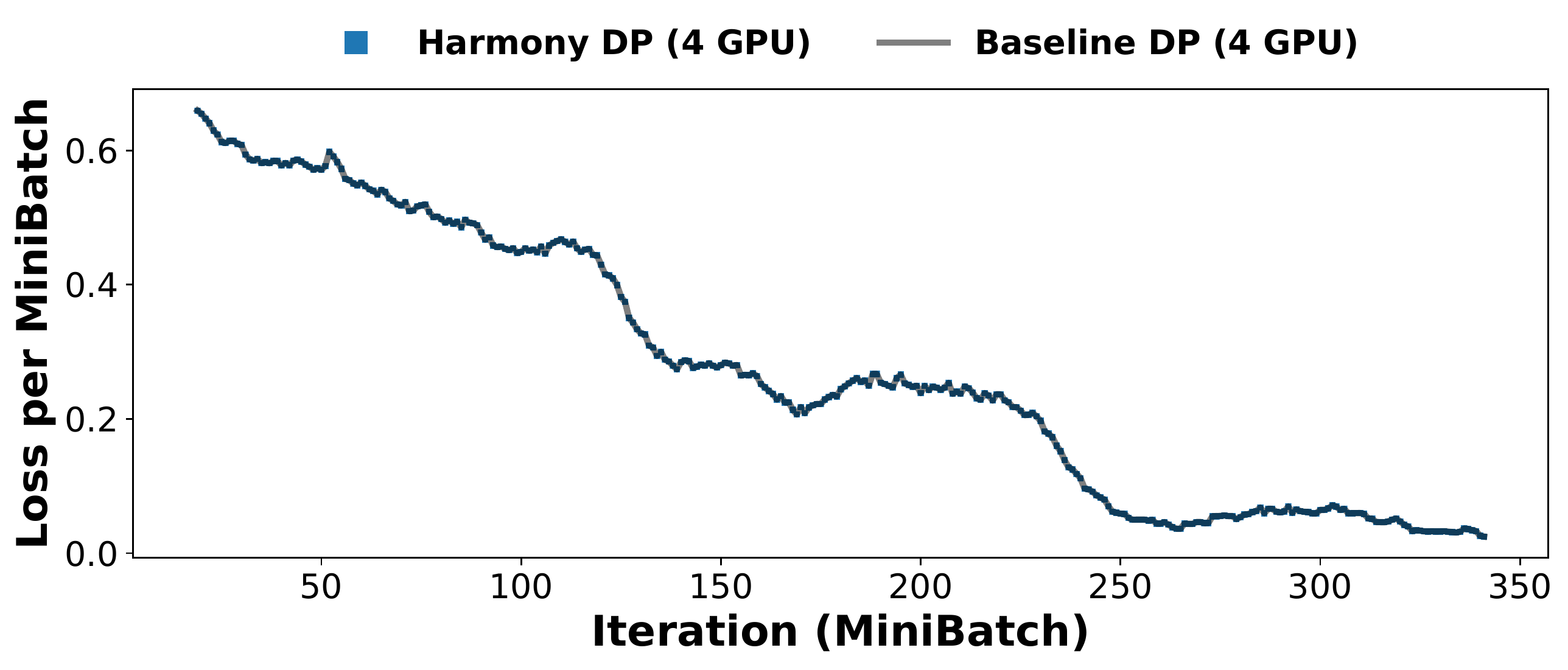}
      \vspace{-4ex}
      \caption{\sysname vs. data-parallelism baseline.} 
    \end{subfigure}
\vspace{-3ex}
\caption{Correctness of \sysname. An example of fine-tuning BERT-Large on MRPC of GLUE with reported hyper-parameters~\cite{devlin2018bert} and baseline code~\cite{PTBERT}. 
\sysname matches the baseline exactly for every minibatch.}
\label{fig:curve_bert}
\vspace{-1ex}
\end{figure}
\subsection{Correctness of Training in \sysname}
\label{sec:correctness}
\sysname{} provides synchronous SGD semantics and should leave convergence properties of models unchanged compared to settings where the entire model would fit in memory.  To validate this, we compare the training loss for every minibatch in \sysname{} (with swaps) with the equivalent training loss of a baseline scheme without swaps, when using the same hyper-parameters and for models that can fit in GPU memory.  \sysname{} PP provides a single-GPU abstraction, and hence we compare it to accuracy results from single-GPU runs.
In Figure~\ref{fig:curve_bert}, fine-tuning results of BERT-Large on downstream MRPC tasks show a \emp{perfect match in loss values for every minibatch} between \sysname{}'s schemes and baseline runs.
We also achieve \textit{perfect match in the final evaluation accuracy} of the trained model: 88.0\% across \sysname{} and baseline runs.

\begin{figure}[!t]
  \centering
  \includegraphics[width=\linewidth]{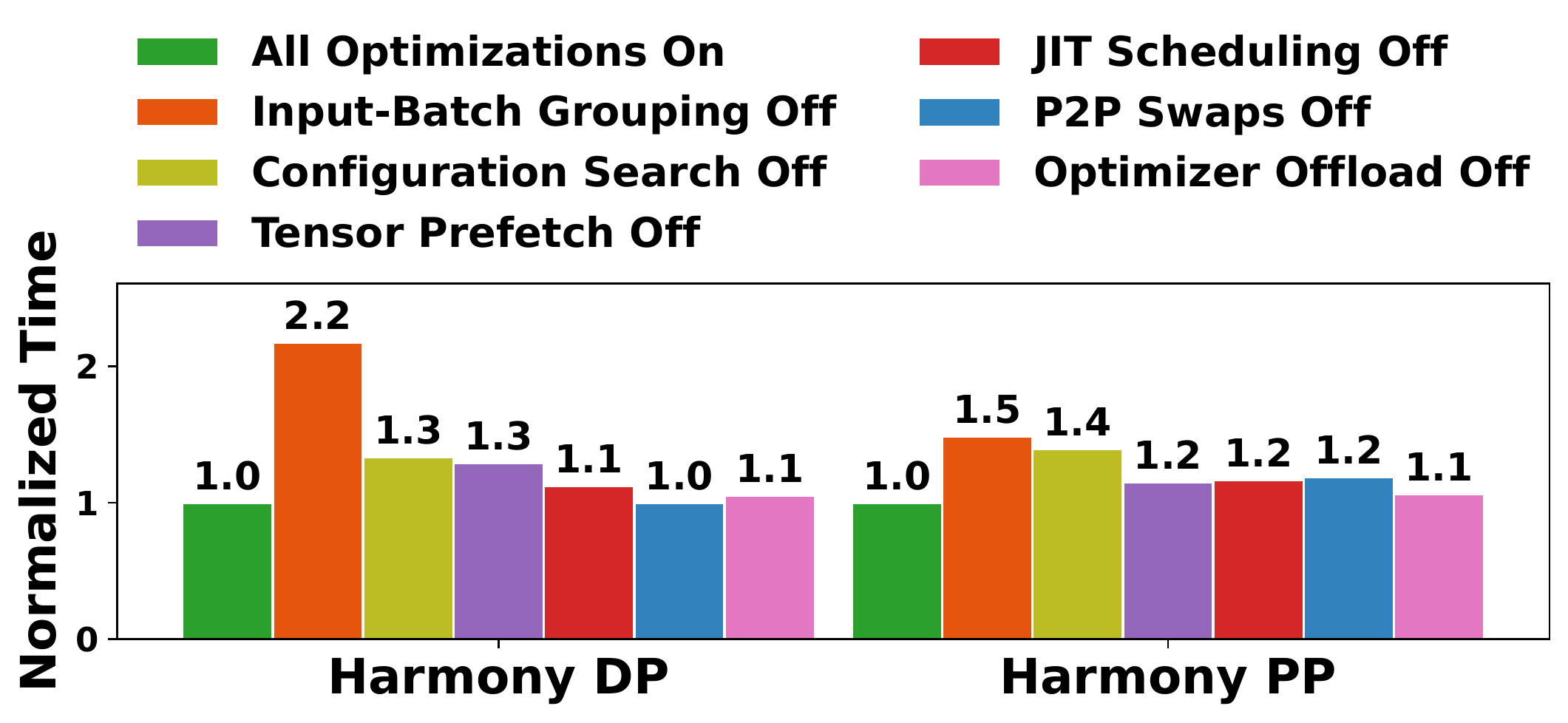}
  \vspace{-5ex}
  \caption{Efficiency breakdown of \sysname for training GPT2 on 4 GPUs. Each bar shows the resulting slowdown when turning off only one optimization while keep others on. Y-axis is normalized against ``All Optimizations On'' for \sysname DP and PP separately. Higher is worse.} 
  \label{fig:breakdown}
\end{figure}

\subsection{Efficiency Breakdown of \sysname}
\label{sec:breakdown_of_optimizations}
Figure~\ref{fig:breakdown} analyzes the efficacy of \system{}'s optimizations.
We highlight five key takeaways. 
First, \textit{input-batch grouping} significantly reduces swap load and increases arithmetic intensity; without this optimization iteration times are $2.2\times$ and $1.5\times$ slower in \sysname{} DP and PP respectively.  
Second, expert (manually) picked layer packs and microbatch sizes for even repeated-structure transformer-based DNNs result in $1.3$--$1.4\times$ worse throughput compared to \sysname{}'s automated \textit{configuration search}.
Third, forgoing \textit{tensor prefetch} can result in up to $1.3\times$ slower iteration times.
Fourth, excluding \textit{jit scheduling} and \textit{optimizer offload} can individually degrade throughput by up to $1.2\times$, although \textit{optimizer offloading} seems to be less critical.
Fifth, \textit{p2p swaps} don't provide any benefits for \sysname DP, but \sysname{} PP which actively uses GPU-GPU swaps across layer packs in the pipeline can suffer degraded iteration times by as much as $1.2\times$ when disabling \textit{p2p swaps}.

\begin{table}[!t]
\small
\center
\caption{Configuration search results and Scheduler end-to-end time (config search, task graph generation, runtime estimation) with \sysname PP (4 GPUs, minibatch size 64).}
\label{tab:search_time}
\vspace{-2ex}
\begin{tabular}{@{}ccccc@{}}
\toprule
Model    & BERT96 & GPT2  & VGG416 & ResNet1K \\ \midrule
$U_F$    & 16     & 4     & 8      & 32       \\ 
$|P_F|$  & 24     & 10    & 15     & 2        \\ 
$U_B$    & 16     & 4     & 8      & 32       \\ 
$|P_B|$  & 25     & 17    & 16     & 9        \\ \midrule
Time (s) & 1.4    & 0.7   & 17.7   & 31.6     \\ \bottomrule
\end{tabular}
\vspace{-1ex}
\end{table}

\begin{figure}[!t]
  \centering
  \includegraphics[width=0.65\linewidth]{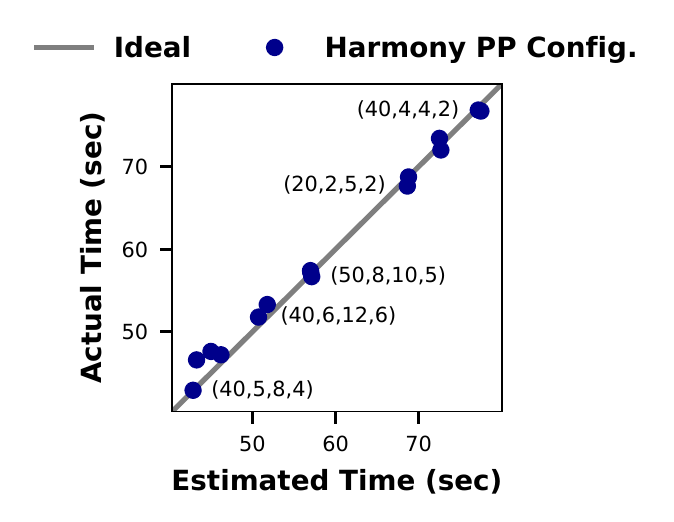}
  \vspace{-3ex}
  \caption{Accuracy of \sysname{}'s Runtime Estimator. We compare estimated iteration time with actual time for training BERT-Large with a mini-batch size of 600 on 4 GPUs using \sysname{} PP. Each dot represents a Harmony configuration $(U_F, |P_F|, U_B, |P_B|)$ and the 15 points here are sampled randomly from all the configurations that \sysname{} explores. The relative difference between estimated and actual time is within 5\% on average.}
  \label{fig:est_acc}
\end{figure}

\subsection{Scheduler and Configuration Search}
\label{sec:scheduler_effectiveness}
To evaluate the effectiveness of \sysname{} Scheduler, we measure its end-to-end time including iterative configuration search, task graph generation, and runtime estimation, until the best configuration is selected. 
Table~\ref{tab:search_time} shows that \emp{reaching the best configuration takes at most 32 seconds}.
For transformers, it is about 1 second;
CNNs like ResNet1K are much deeper and richer in diversity of layer attributes (memory size and compute time).

Figure~\ref{fig:est_acc} evaluates the quality of \sysname{}'s \textit{Runtime Estimator}. 
It compares the estimated training time with actual training time in each searched configuration. 
Estimated training time is obtained from \sysname{}'s \textit{event-driven simulator} (\secref{sec:config_search}); for each configuration, the simulator uses a \sysname{} task graph and layer profiles for estimating end-to-end iteration time.
We observe that \textit{\sysname{}'s estimates are accurate}, giving us confidence in its selection of configurations with the best throughput. 

\begin{figure*}[!t]
  \centering
  \includegraphics[width=0.85\linewidth]{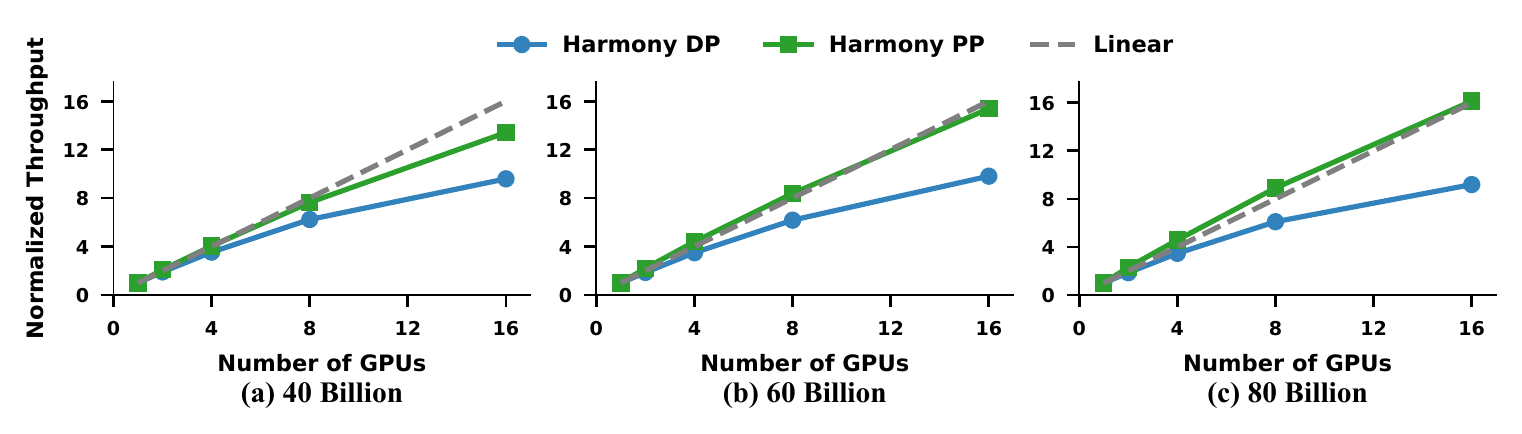}
  \vspace{-4ex}
  \caption{Scalability of \sysname in training massive models of 10s of billions of parameters with 16 V100s on a DGX-2 server. The 80-Billion model saturates CPU memory capacity (1.5TB). Throughput is normalized against single-GPU \sysname DP.} 
  \label{fig:scalability_dgx}
\end{figure*}
\begin{figure}[!t]
  \centering
  \includegraphics[width=\linewidth]{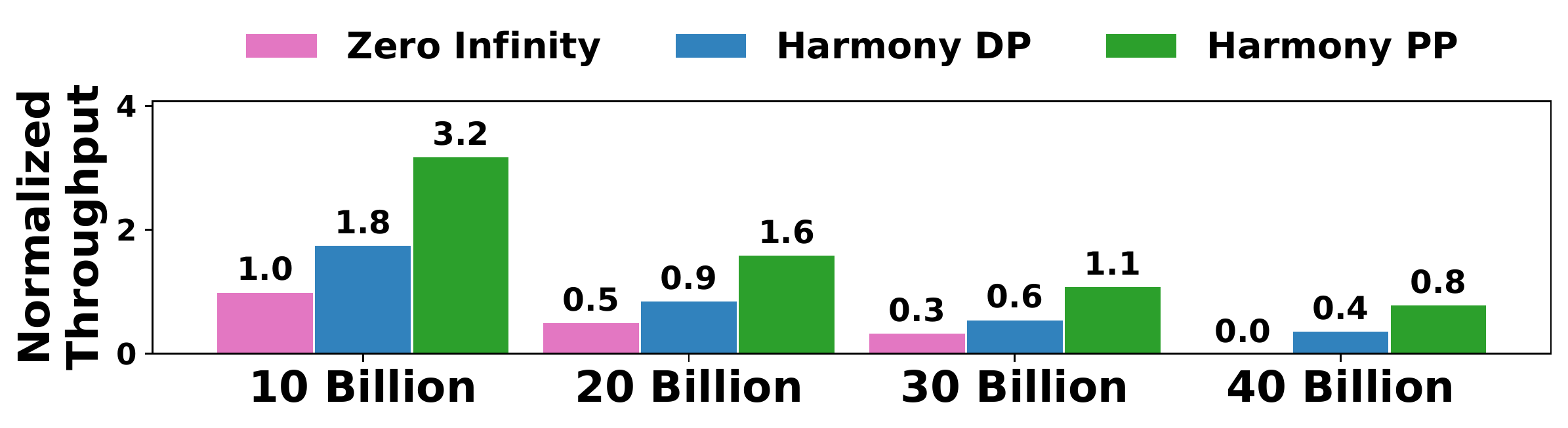}
  \vspace{-5ex}
  \caption{Training massive models of 10s of billions of parameters at the limit of single-server CPU memory capacity (750GB CPU and eight GTX1080Tis). Throughput is normalized against \zinf's 10-Billion model.} 
  \label{fig:two_numa_largest_models}
\end{figure}

\subsection{Scaling Model Size and Number of GPUs}
\label{sec:scalability_1080ti}
\label{sec:scalability_dgx2}
To evaluate how \sysname{} scales, we now use two beefier servers as mentioned in \secref{sec:eval_setup} -- i) a server with \textit{eight} GTX-1080Tis (11GB) and ii) a DGX-2 with 16 V100 GPUs (32GB).
We use this setup to understand not only the limits of how large a model \sysname{} can train given CPU memory capacity bounds but also \sysname{}'s scalability in number of GPUs.
We customize the GPT2 model to scale up to tens of billions of parameters~\cite{rajbhandari2021zeroinf}.

First, we study the limit of trainable model size.
Figure~\ref{fig:two_numa_largest_models} shows the throughput of training such models on an 8$\times$ GTX-1080Ti server.
For fairness, \zinf shares the same configuration as \sysname and with all optimization flags on.
For the 10$\sim$30-billion models, \emp{\sysname DP and PP are still consistently faster than \zinf by up to $1.8\times$ and $3.2\times$}, respectively.
At 40 billion parameters, the working set size of the model hits the limits of the server's CPU memory capacity; \textit{\sysname{} offers proportionally scaled throughput} compared to the 10-billion parameter model on account of $4\times$ more compute, while \zinf fails to train due to running out of CPU memory.

\begin{figure}[!t]
  \centering
  \includegraphics[width=\linewidth]{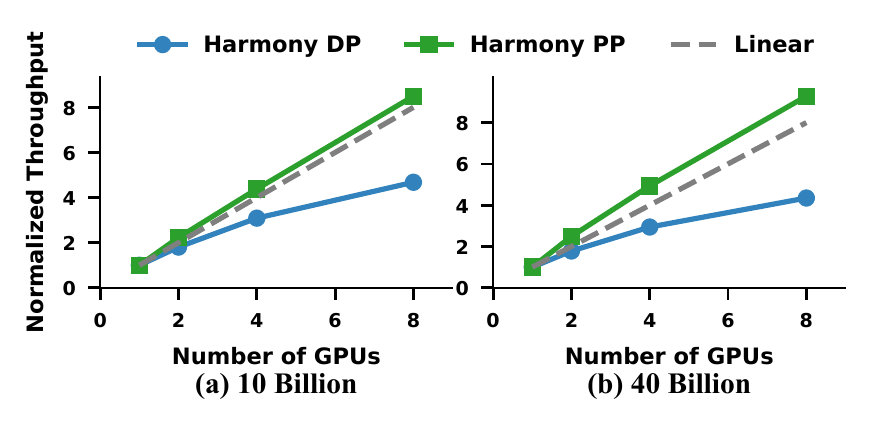}
  \vspace{-6ex}
  \caption{Scalability of \sysname in training massive models of 10s of billions of parameters with eight GTX1080Tis. Throughput is normalized against single-GPU \sysname DP.} 
  \label{fig:scalability_largest_models_onecol}
\end{figure}

Second, we scale the number of GPUs from 1 to 8 on the GTX-1080Ti server (Figure~\ref{fig:scalability_largest_models_onecol}) and from 1 to 16 on the DGX-2 server (Figure~\ref{fig:scalability_dgx}).
When using \emp{\system{} PP, training throughput scales linearly} due to reduced swap overhead and advantages of p2p swaps.
\system{} DP, despite performing well, is affected by the overhead of weight swaps duplicated across GPUs, and the performance gap between \system{} DP and PP widens as model size grows (due to greater swap volume).

\section{Discussion}
\label{sec:discussion}
\niparagraph{Scope of \sysname{}.}
\sysname{} aims at training massive models that are \textit{out of GPU memory capacity}, which is of great value for \textit{developing}, \textit{debugging}, and \textit{fine-tuning} massive models on a single commodity server with only several GPUs~\cite{eliad2021ftpipe}. 
Our scope differs from prior work that requires memory footprint to \textit{fit GPU memory capacity} and focuses on \textit{pre-training} with hundreds of GPUs in a datacenter~\cite{narayanan2020memory,fan2021dapple,shoeybi2019megatron,rajbhandari2019zero}.
Pre-training extremely large models on a commodity server might be infeasible.
For instance, pre-training GPT3 end to end requires 314 ZettaFLOPs~\cite{brown2020language} and takes several months of training even with thousands of cutting-edge GPUs~\cite{narayanan2021efficient}.
There is no denying that training on a large cluster will naturally result in speedier training. 
However, despite this limitation, we believe that \sysname{} can still enable development and debugging of such models on modest deployments (before they are deployed for pre-training at a large scale), and fine-tuning of massive models that requires less than 10s of exaFLOPs~\cite{kaplan2020scaling,devlin2018bert,bhandare2020finetuning} clocking in at days with a commodity server~\cite{huggingface2021finetuning}.

\niparagraph{Multi-machine Training.}
\sysname{}'s prototype operates on a single machine.
However, \textit{task decomposition} and \textit{late binding}, together with \sysname{}'s four key optimizations 
and \wap, all extend to multi-machine training.
When collective capacity of all GPUs is sufficient to hold the memory footprint of massive models, memory swapping becomes unnecessary. 
Nonetheless, the single-GPU abstraction of \sysname{} can still benefit developers by decoupling the model definition from a particular training parallelism, allowing to focus on model development without worrying about complexity and deadlock~\cite{narayanan2019pipedream,lepikhin2021gshard} in parallel training. 

\section{Conclusion}
\label{sec:conclusion}
One of the main challenges for training massive DNN models on single-server multi-GPU deployments is the limited GPU memory capacity.
Current solutions that rely on virtualizing GPU memory incur excessive swap overheads.
We advocate rethinking how DNN frameworks schedule computation and move data, and we articulate the principles, functionality, and optimizations needed to push the boundaries of training massive models efficiently on modest deployments.
Across various massive DNN models, \sysname is able to reduce swap load by up to two orders of magnitude and obtain a training throughput speedup of up to 7.6$\times$ over highly optimized baselines with virtualized memory.

\clearpage
\bibliographystyle{ACM-Reference-Format}
\bibliography{references}

\clearpage
\appendix
\section*{Appendix}

\section{NP-Hardness Proof}
In this section, we prove that \textit{finding the optimal layer packs that minimize training time is NP-hard}.
To that end, we introduce a simplified version of the scheduling (pack selection) problem arising in \sysname, for which we will prove NP-hardness.
\begin{definition}
    In the \textbf{\sysname scheduling problem}, we are given as input:
    \begin{itemize}
        \item $B$: number of microbatches in a batch,
        \item $G$: number of GPUs,
        \item $M$: memory per GPU,
        \item a sequence of $n$ layers, each with processing time $p_i$ and size of weights $m_i$.
    \end{itemize}
    A solution to the problem is a partitioning of the set of layers into contiguous intervals called \emph{packs}.
    A solution is feasible if for every pack $S \subseteq \{1,...,n\}$, the layer weights fit on a GPU: $\sum_{i \in S} m_i \le M$.
    The cost, or \emph{makespan}, of a solution is the time it takes to execute all packs, with a round-robin assignment of packs to GPUs, on $B$ microbatches.
    More precisely, each pack in order is scheduled as follows:
    \begin{itemize}
        \item the $j$-th pack, $S_j$, is executed on GPU number $1 + ((j-1) \mod G)$ (i.e., round-robin),
        \item for each microbatch $b=1,...,B$, it starts executing on that GPU at the earliest time when that GPU is idle and (if $j > 1$) the $b$-th microbatch is done executing on pack $S_{j-1}$ (on the previous GPU),
        \item the execution of that microbatch takes time $\sum_{i \in S_j} p_i$.
    \end{itemize}
    See Figure~\ref{fig:problem} for an illustration.
\end{definition}
In the decision version of the above problem, we are additionally given a target makespan $T$ and we need to determine whether there is a packing whose makespan is at most $T$. 

Note that the following aspects of the real \sysname scheduling problem are missing from our simplified version (which would make it even harder if they were present):
\begin{itemize}
    \item backward pass and weight updates (alternatively, one could think of the simplified version as a special case of the forward+backward pass case where the forward processing time of each layer is 0 and the order of layers is reversed), 
    \item data transfer latencies (p2p swaps, CPU-GPU swaps, etc.), 
    \item memory usage of any non-weight tensors (like activations),
    \item in addition, in our reduction we will only use $G=2$ and $B=3$.
\end{itemize}

\begin{proposition}
    \textbf{The \sysname scheduling problem is NP-hard.}
\end{proposition}

\begin{proof}
    We reduce from the well-known NP-hard Partition problem.
    An input to the Partition problem is a collection $a_1, ..., a_n \in \mathbb{Z}_+$ of positive integers,
    and the task is to determine whether this collection can be partitioned into two subsets of equal sum.
    
    Given an instance of Partition,
    we produce an instance of the \sysname scheduling problem as follows.
    We set $B=3$, $G=2$, and $M=7$.
    We produce $3n+4$ layers, as follows; see also Table~\ref{tab:reduction_layers}.
    Let $A$ be a large processing time quantity, say $A = 6 \sum_{i=1}^n a_i$.
    The first two layers will have $p_1 = p_2 = 8A$ and $m_1 = m_2 = 6$.
    The last two will be the same: $p_{3n+3} = p_{3n+4} = 8A$ and $m_{3n+3} = m_{3n+4} = 6$.
    For every number $a_i$, $i=1,...,n$ from the Partition instance,
    we generate three layers: $p_{3i} = p_{3i+2} = 5A$, $p_{3i+1} = a_i$, $m_{3i} = m_{3i+2} = 4$, $m_{3i+1} = 2$.
    
    \begin{table}[!t]
    \small
    \center
    \caption{The reduction outputs the following set of layers for a Partition instance $a_1, ..., a_n$, where $A$ is some large enough time.}
    \label{tab:reduction_layers}
    \vspace{-2ex}
    \begin{tabular}{@{}ccc@{}}
    \toprule
    \textbf{index $j$} & \textbf{time $p_j$} & \textbf{size $m_j$} \\ \midrule
    1 & 8A & 6 \\
    2 & 8A & 6 \\
    \midrule
    \multicolumn{3}{c}{for $i=1,...,n$:}\\
    $3i$ & 5A & 4 \\
    $3i+1$ & $a_i$ & 2 \\
    $3i+2$ & 5A & 4 \\
    \midrule
    $3n+3$ & 8A & 6 \\
    $3n+4$ & 8A & 6 \\
    \bottomrule
    \end{tabular}
    \vspace{-3ex}
    \end{table}

    Let us first informally explain how this instance models the Partition instance.
    The first two and the last two layers are so large that they must be in single-layer packs.
    The only way that non-trivial packs can be formed is that for each $i$, the layer $3i+1$ can be packed together with $3i$ or with $3i+2$. This corresponds to picking the number $a_i$ for the first or for the second subset of the partition. (The layer $3i+1$ can also be placed in a pack by itself, but this leads to high makespan.)
    
    We begin with the following observation:
    at the time when microbatch 1 of layer 1 (i.e., pack 1) is being processed, the other GPU is idle,
    and the same holds for
    the last microbatch of the last layer (pack).
    Call these two time periods (each of length $8A$) \emph{forced-idle times}.
    Since the total processing time of all layers (on all microbatches)
    is the same regardless of the partitioning into packs,
    the makespan of any solution is at least
    (that total processing time + forced-idle times)/2;
    we define this as the target makespan $T$.
    Further note that $T$ is attained if and only if both GPUs are processing at all times except the forced-idle times.
    Formally, we have
    \begin{align*} T &= \frac{B \cdot \sum_{i=1}^{3n+4} p_i + p_1 + p_{3n+4}}{G} \\ &= \frac{3 \cdot \left(2 \cdot 8A + \sum_{i=1}^n \left(5A + a_i + 5A\right) + 2 \cdot 8A \right) + 2 \cdot 8A}{2} \,.
    \end{align*}
    To restate: $T$ is a lower bound on the makespan of any solution, and
    it is attained if and only if GPUs are idle only during forced-idle times.
    
    Now we will show that $a_1, ..., a_n$ is a YES-case of Partition if and only if the \sysname scheduling problem instance output by the reduction admits a solution of makespan $T$.
    
    \newcommand{\tikzbarheight}{60}
\tikzstyle{oddstyle}=[]
\tikzstyle{evenstyle}=[pattern=crosshatch, pattern color=blue!15]

\begin{figure*}[t]
    \centering

    \subfloat[A balanced solution: $a_1 = 6$ goes to GPU 1, while $a_2 = 2$ and $a_3 = 4$ go to GPU 2. By this we mean that the layer $3 \cdot 1 + 1$ (of processing time $6$) is placed together with layer $3 \cdot 1$ in the pack P3, which goes to GPU 1, whereas the layer $3 \cdot 2 + 1$ (of processing time $2$) is placed together with layer $3 \cdot 2 + 2$ in the pack P6, which goes to GPU 2 (and similarly for layer $3 \cdot 3 + 1$, of processing time $4$). 
    GPUs are idle only during forced-idle times, and optimal makespan is achieved. With $A$ large enough, our construction guarantees that no GPU needs to wait for the other GPU (except for the forced-idle time at the beginning).]{
        \begin{tikzpicture}[scale=0.3, every text node part/.style={align=center, font=\small}]  

			\draw[oddstyle] (0,0*\tikzbarheight) rectangle (4,-0.05*\tikzbarheight) node[midway] {1 \\ P1};
			\draw [decorate,decoration={brace,amplitude=5pt,mirror,raise=4pt}] (0,-0.1*\tikzbarheight) -- (4,-0.1*\tikzbarheight) node [midway,yshift=-0.5cm] {forced-idle time};
			\draw[oddstyle] (4,0*\tikzbarheight) rectangle (8,-0.05*\tikzbarheight) node[midway] {2 \\ P1};
			\draw[oddstyle] (8,0*\tikzbarheight) rectangle (12,-0.05*\tikzbarheight) node[midway] {3 \\ P1};
			\draw[oddstyle] (4,-0.05*\tikzbarheight) rectangle (8,-0.1*\tikzbarheight) node[midway] {1 \\ P2};
			\draw[oddstyle] (8,-0.05*\tikzbarheight) rectangle (12,-0.1*\tikzbarheight) node[midway] {2 \\ P2};
			\draw[oddstyle] (12,-0.05*\tikzbarheight) rectangle (16,-0.1*\tikzbarheight) node[midway] {3 \\ P2};
			\draw[evenstyle] (12,0*\tikzbarheight) rectangle (14.8,-0.05*\tikzbarheight) node[midway] {1 \\ P3};
			\draw[evenstyle] (14.8,0*\tikzbarheight) rectangle (17.6,-0.05*\tikzbarheight) node[midway] {2 \\ P3};
			\draw[evenstyle] (17.6,0*\tikzbarheight) rectangle (20.4,-0.05*\tikzbarheight) node[midway] {3 \\ P3};
			\draw[evenstyle] (16,-0.05*\tikzbarheight) rectangle (18.5,-0.1*\tikzbarheight) node[midway] {1 \\ P4};
			\draw[evenstyle] (18.5,-0.05*\tikzbarheight) rectangle (21,-0.1*\tikzbarheight) node[midway] {2 \\ P4};
			\draw[evenstyle] (21,-0.05*\tikzbarheight) rectangle (23.5,-0.1*\tikzbarheight) node[midway] {3 \\ P4};
			\draw[oddstyle] (20.4,0*\tikzbarheight) rectangle (22.9,-0.05*\tikzbarheight) node[midway] {1 \\ P5};
			\draw[oddstyle] (22.9,0*\tikzbarheight) rectangle (25.4,-0.05*\tikzbarheight) node[midway] {2 \\ P5};
			\draw[oddstyle] (25.4,0*\tikzbarheight) rectangle (27.9,-0.05*\tikzbarheight) node[midway] {3 \\ P5};
			\draw[oddstyle] (23.5,-0.05*\tikzbarheight) rectangle (26.1,-0.1*\tikzbarheight) node[midway] {1 \\ P6};
			\draw[oddstyle] (26.1,-0.05*\tikzbarheight) rectangle (28.7,-0.1*\tikzbarheight) node[midway] {2 \\ P6};
			\draw[oddstyle] (28.7,-0.05*\tikzbarheight) rectangle (31.3,-0.1*\tikzbarheight) node[midway] {3 \\ P6};
			\draw[evenstyle] (27.9,0*\tikzbarheight) rectangle (30.4,-0.05*\tikzbarheight) node[midway] {1 \\ P7};
			\draw[evenstyle] (30.4,0*\tikzbarheight) rectangle (32.9,-0.05*\tikzbarheight) node[midway] {2 \\ P7};
			\draw[evenstyle] (32.9,0*\tikzbarheight) rectangle (35.4,-0.05*\tikzbarheight) node[midway] {3 \\ P7};
			\draw[evenstyle] (31.3,-0.05*\tikzbarheight) rectangle (34,-0.1*\tikzbarheight) node[midway] {1 \\ P8};
			\draw[evenstyle] (34,-0.05*\tikzbarheight) rectangle (36.7,-0.1*\tikzbarheight) node[midway] {2 \\ P8};
			\draw[evenstyle] (36.7,-0.05*\tikzbarheight) rectangle (39.4,-0.1*\tikzbarheight) node[midway] {3 \\ P8};
			\draw[oddstyle] (35.4,0*\tikzbarheight) rectangle (39.4,-0.05*\tikzbarheight) node[midway] {1 \\ P9};
			\draw[oddstyle] (39.4,0*\tikzbarheight) rectangle (43.4,-0.05*\tikzbarheight) node[midway] {2 \\ P9};
			\draw[oddstyle] (43.4,0*\tikzbarheight) rectangle (47.4,-0.05*\tikzbarheight) node[midway] {3 \\ P9};
			\draw[oddstyle] (39.4,-0.05*\tikzbarheight) rectangle (43.4,-0.1*\tikzbarheight) node[midway] {1 \\ P10};
			\draw[oddstyle] (43.4,-0.05*\tikzbarheight) rectangle (47.4,-0.1*\tikzbarheight) node[midway] {2 \\ P10};
			\draw[oddstyle] (47.4,-0.05*\tikzbarheight) rectangle (51.4,-0.1*\tikzbarheight) node[midway] {3 \\ P10};
			\draw [decorate,decoration={brace,amplitude=5pt,mirror,raise=4pt}] (47.4,-0.1*\tikzbarheight) -- (51.4,-0.1*\tikzbarheight) node [midway,yshift=-0.5cm] {forced-idle time};

        \end{tikzpicture}
    \label{fig_a}}
    \vspace{2.5em}

    \subfloat[The fourth layer ($3i+1$ for $i=1$) is put in a pack (P4) of its own, which immediately causes unforced idle times to appear. This solution to the Harmony scheduling problem instance does not correspond to any solution of the Partition instance.]{
        \begin{tikzpicture}[scale=0.3, every text node part/.style={align=center, font=\small}]  

			\draw[oddstyle] (0,0*\tikzbarheight) rectangle (4,-0.05*\tikzbarheight) node[midway] {1 \\ P1};
			\draw [decorate,decoration={brace,amplitude=5pt,mirror,raise=4pt}] (0,-0.1*\tikzbarheight) -- (4,-0.1*\tikzbarheight) node [midway,yshift=-0.5cm] {forced-idle time};
			\draw[oddstyle] (4,0*\tikzbarheight) rectangle (8,-0.05*\tikzbarheight) node[midway] {2 \\ P1};
			\draw[oddstyle] (8,0*\tikzbarheight) rectangle (12,-0.05*\tikzbarheight) node[midway] {3 \\ P1};
			\draw[oddstyle] (4,-0.05*\tikzbarheight) rectangle (8,-0.1*\tikzbarheight) node[midway] {1 \\ P2};
			\draw[oddstyle] (8,-0.05*\tikzbarheight) rectangle (12,-0.1*\tikzbarheight) node[midway] {2 \\ P2};
			\draw[oddstyle] (12,-0.05*\tikzbarheight) rectangle (16,-0.1*\tikzbarheight) node[midway] {3 \\ P2};
			\draw[evenstyle] (12,0*\tikzbarheight) rectangle (14.5,-0.05*\tikzbarheight) node[midway] {1 \\ P3};
			\draw[evenstyle] (14.5,0*\tikzbarheight) rectangle (17,-0.05*\tikzbarheight) node[midway] {2 \\ P3};
			\draw[evenstyle] (17,0*\tikzbarheight) rectangle (19.5,-0.05*\tikzbarheight) node[midway] {3 \\ P3};
			\draw[evenstyle] (16,-0.05*\tikzbarheight) rectangle (16.3,-0.1*\tikzbarheight) node[midway] {1 \\ P4};
			\draw[evenstyle] (17,-0.05*\tikzbarheight) rectangle (17.3,-0.1*\tikzbarheight) node[midway] {2 \\ P4};
			\draw[evenstyle] (19.5,-0.05*\tikzbarheight) rectangle (19.8,-0.1*\tikzbarheight) node[midway] {3 \\ P4};
			\draw[oddstyle] (19.5,0*\tikzbarheight) rectangle (22,-0.05*\tikzbarheight) node[midway] {1 \\ P5};
			\draw[oddstyle] (22,0*\tikzbarheight) rectangle (24.5,-0.05*\tikzbarheight) node[midway] {2 \\ P5};
			\draw[oddstyle] (24.5,0*\tikzbarheight) rectangle (27,-0.05*\tikzbarheight) node[midway] {3 \\ P5};
			\draw[oddstyle] (22,-0.05*\tikzbarheight) rectangle (24.5,-0.1*\tikzbarheight) node[midway] {1 \\ P6};
			\draw[oddstyle] (24.5,-0.05*\tikzbarheight) rectangle (27,-0.1*\tikzbarheight) node[midway] {2 \\ P6};
			\draw[oddstyle] (27,-0.05*\tikzbarheight) rectangle (29.5,-0.1*\tikzbarheight) node[midway] {3 \\ P6};
			\draw[evenstyle] (27,0*\tikzbarheight) rectangle (29.6,-0.05*\tikzbarheight) node[midway] {1 \\ P7};
			\draw[evenstyle] (29.6,0*\tikzbarheight) rectangle (32.2,-0.05*\tikzbarheight) node[midway] {2 \\ P7};
			\draw[evenstyle] (32.2,0*\tikzbarheight) rectangle (34.8,-0.05*\tikzbarheight) node[midway] {3 \\ P7};
			\draw[evenstyle] (29.6,-0.05*\tikzbarheight) rectangle (32.1,-0.1*\tikzbarheight) node[midway] {1 \\ P8};
			\draw[evenstyle] (32.2,-0.05*\tikzbarheight) rectangle (34.7,-0.1*\tikzbarheight) node[midway] {2 \\ P8};
			\draw[evenstyle] (34.8,-0.05*\tikzbarheight) rectangle (37.3,-0.1*\tikzbarheight) node[midway] {3 \\ P8};
			\draw[oddstyle] (34.8,0*\tikzbarheight) rectangle (37.5,-0.05*\tikzbarheight) node[midway] {1 \\ P9};
			\draw[oddstyle] (37.5,0*\tikzbarheight) rectangle (40.2,-0.05*\tikzbarheight) node[midway] {2 \\ P9};
			\draw[oddstyle] (40.2,0*\tikzbarheight) rectangle (42.9,-0.05*\tikzbarheight) node[midway] {3 \\ P9};
			\draw[oddstyle] (37.5,-0.05*\tikzbarheight) rectangle (41.5,-0.1*\tikzbarheight) node[midway] {1 \\ P10};
			\draw[oddstyle] (41.5,-0.05*\tikzbarheight) rectangle (45.5,-0.1*\tikzbarheight) node[midway] {2 \\ P10};
			\draw[oddstyle] (45.5,-0.05*\tikzbarheight) rectangle (49.5,-0.1*\tikzbarheight) node[midway] {3 \\ P10};
			\draw[evenstyle] (42.9,0*\tikzbarheight) rectangle (46.9,-0.05*\tikzbarheight) node[midway] {1 \\ P11};
			\draw[evenstyle] (46.9,0*\tikzbarheight) rectangle (50.9,-0.05*\tikzbarheight) node[midway] {2 \\ P11};
			\draw[evenstyle] (50.9,0*\tikzbarheight) rectangle (54.9,-0.05*\tikzbarheight) node[midway] {3 \\ P11};
			\draw [decorate,decoration={brace,amplitude=5pt,mirror,raise=4pt}] (50.9,-0.1*\tikzbarheight) -- (54.9,-0.1*\tikzbarheight) node [midway,yshift=-0.5cm] {forced-idle time};
			
		\end{tikzpicture}
    \label{fig_b}}
    \vspace{2.5em}

    \subfloat[Unbalanced solution: $a_1 = 6$ and $a_2 = 2$ go to GPU 1, while $a_3 = 4$ goes to GPU 2. Thus GPU 1 has more load, and an unforced idle time appears after P8 for GPU 2.]{
        \begin{tikzpicture}[scale=0.3, every text node part/.style={align=center, font=\small}]  

			\draw[oddstyle] (0,0*\tikzbarheight) rectangle (4,-0.05*\tikzbarheight) node[midway] {1 \\ P1};
			\draw [decorate,decoration={brace,amplitude=5pt,mirror,raise=4pt}] (0,-0.1*\tikzbarheight) -- (4,-0.1*\tikzbarheight) node [midway,yshift=-0.5cm] {forced-idle time};
			\draw[oddstyle] (4,0*\tikzbarheight) rectangle (8,-0.05*\tikzbarheight) node[midway] {2 \\ P1};
			\draw[oddstyle] (8,0*\tikzbarheight) rectangle (12,-0.05*\tikzbarheight) node[midway] {3 \\ P1};
			\draw[oddstyle] (4,-0.05*\tikzbarheight) rectangle (8,-0.1*\tikzbarheight) node[midway] {1 \\ P2};
			\draw[oddstyle] (8,-0.05*\tikzbarheight) rectangle (12,-0.1*\tikzbarheight) node[midway] {2 \\ P2};
			\draw[oddstyle] (12,-0.05*\tikzbarheight) rectangle (16,-0.1*\tikzbarheight) node[midway] {3 \\ P2};
			\draw[evenstyle] (12,0*\tikzbarheight) rectangle (14.8,-0.05*\tikzbarheight) node[midway] {1 \\ P3};
			\draw[evenstyle] (14.8,0*\tikzbarheight) rectangle (17.6,-0.05*\tikzbarheight) node[midway] {2 \\ P3};
			\draw[evenstyle] (17.6,0*\tikzbarheight) rectangle (20.4,-0.05*\tikzbarheight) node[midway] {3 \\ P3};
			\draw[evenstyle] (16,-0.05*\tikzbarheight) rectangle (18.5,-0.1*\tikzbarheight) node[midway] {1 \\ P4};
			\draw[evenstyle] (18.5,-0.05*\tikzbarheight) rectangle (21,-0.1*\tikzbarheight) node[midway] {2 \\ P4};
			\draw[evenstyle] (21,-0.05*\tikzbarheight) rectangle (23.5,-0.1*\tikzbarheight) node[midway] {3 \\ P4};
			\draw[oddstyle] (20.4,0*\tikzbarheight) rectangle (23,-0.05*\tikzbarheight) node[midway] {1 \\ P5};
			\draw[oddstyle] (23,0*\tikzbarheight) rectangle (25.6,-0.05*\tikzbarheight) node[midway] {2 \\ P5};
			\draw[oddstyle] (25.6,0*\tikzbarheight) rectangle (28.2,-0.05*\tikzbarheight) node[midway] {3 \\ P5};
			\draw[oddstyle] (23.5,-0.05*\tikzbarheight) rectangle (26,-0.1*\tikzbarheight) node[midway] {1 \\ P6};
			\draw[oddstyle] (26,-0.05*\tikzbarheight) rectangle (28.5,-0.1*\tikzbarheight) node[midway] {2 \\ P6};
			\draw[oddstyle] (28.5,-0.05*\tikzbarheight) rectangle (31,-0.1*\tikzbarheight) node[midway] {3 \\ P6};
			\draw[evenstyle] (28.2,0*\tikzbarheight) rectangle (30.7,-0.05*\tikzbarheight) node[midway] {1 \\ P7};
			\draw[evenstyle] (30.7,0*\tikzbarheight) rectangle (33.2,-0.05*\tikzbarheight) node[midway] {2 \\ P7};
			\draw[evenstyle] (33.2,0*\tikzbarheight) rectangle (35.7,-0.05*\tikzbarheight) node[midway] {3 \\ P7};
			\draw[evenstyle] (31,-0.05*\tikzbarheight) rectangle (33.7,-0.1*\tikzbarheight) node[midway] {1 \\ P8};
			\draw[evenstyle] (33.7,-0.05*\tikzbarheight) rectangle (36.4,-0.1*\tikzbarheight) node[midway] {2 \\ P8};
			\draw[evenstyle] (36.4,-0.05*\tikzbarheight) rectangle (39.1,-0.1*\tikzbarheight) node[midway] {3 \\ P8};
			\draw[oddstyle] (35.7,0*\tikzbarheight) rectangle (39.7,-0.05*\tikzbarheight) node[midway] {1 \\ P9};
			\draw[oddstyle] (39.7,0*\tikzbarheight) rectangle (43.7,-0.05*\tikzbarheight) node[midway] {2 \\ P9};
			\draw[oddstyle] (43.7,0*\tikzbarheight) rectangle (47.7,-0.05*\tikzbarheight) node[midway] {3 \\ P9};
			\draw[oddstyle] (39.7,-0.05*\tikzbarheight) rectangle (43.7,-0.1*\tikzbarheight) node[midway] {1 \\ P10};
			\draw[oddstyle] (43.7,-0.05*\tikzbarheight) rectangle (47.7,-0.1*\tikzbarheight) node[midway] {2 \\ P10};
			\draw[oddstyle] (47.7,-0.05*\tikzbarheight) rectangle (51.7,-0.1*\tikzbarheight) node[midway] {3 \\ P10};
			\draw [decorate,decoration={brace,amplitude=5pt,mirror,raise=4pt}] (47.7,-0.1*\tikzbarheight) -- (51.7,-0.1*\tikzbarheight) node [midway,yshift=-0.5cm] {forced-idle time};

        \end{tikzpicture}
    \label{fig_c}}
    \vspace{2.5em}

    \subfloat[Unbalanced solution: $a_1 = 6$ and $a_2 = 2$ go to GPU 2, while $a_3 = 4$ goes to GPU 1. Thus GPU 2 has more load, and an unforced idle time appears after P9 for GPU 1.]{
        \begin{tikzpicture}[scale=0.3, every text node part/.style={align=center, font=\small}]  

			\draw[oddstyle] (0,0*\tikzbarheight) rectangle (4,-0.05*\tikzbarheight) node[midway] {1 \\ P1};
			\draw [decorate,decoration={brace,amplitude=5pt,mirror,raise=4pt}] (0,-0.1*\tikzbarheight) -- (4,-0.1*\tikzbarheight) node [midway,yshift=-0.5cm] {forced-idle time};
			\draw[oddstyle] (4,0*\tikzbarheight) rectangle (8,-0.05*\tikzbarheight) node[midway] {2 \\ P1};
			\draw[oddstyle] (8,0*\tikzbarheight) rectangle (12,-0.05*\tikzbarheight) node[midway] {3 \\ P1};
			\draw[oddstyle] (4,-0.05*\tikzbarheight) rectangle (8,-0.1*\tikzbarheight) node[midway] {1 \\ P2};
			\draw[oddstyle] (8,-0.05*\tikzbarheight) rectangle (12,-0.1*\tikzbarheight) node[midway] {2 \\ P2};
			\draw[oddstyle] (12,-0.05*\tikzbarheight) rectangle (16,-0.1*\tikzbarheight) node[midway] {3 \\ P2};
			\draw[evenstyle] (12,0*\tikzbarheight) rectangle (14.5,-0.05*\tikzbarheight) node[midway] {1 \\ P3};
			\draw[evenstyle] (14.5,0*\tikzbarheight) rectangle (17,-0.05*\tikzbarheight) node[midway] {2 \\ P3};
			\draw[evenstyle] (17,0*\tikzbarheight) rectangle (19.5,-0.05*\tikzbarheight) node[midway] {3 \\ P3};
			\draw[evenstyle] (16,-0.05*\tikzbarheight) rectangle (18.8,-0.1*\tikzbarheight) node[midway] {1 \\ P4};
			\draw[evenstyle] (18.8,-0.05*\tikzbarheight) rectangle (21.6,-0.1*\tikzbarheight) node[midway] {2 \\ P4};
			\draw[evenstyle] (21.6,-0.05*\tikzbarheight) rectangle (24.4,-0.1*\tikzbarheight) node[midway] {3 \\ P4};
			\draw[oddstyle] (19.5,0*\tikzbarheight) rectangle (22,-0.05*\tikzbarheight) node[midway] {1 \\ P5};
			\draw[oddstyle] (22,0*\tikzbarheight) rectangle (24.5,-0.05*\tikzbarheight) node[midway] {2 \\ P5};
			\draw[oddstyle] (24.5,0*\tikzbarheight) rectangle (27,-0.05*\tikzbarheight) node[midway] {3 \\ P5};
			\draw[oddstyle] (24.4,-0.05*\tikzbarheight) rectangle (27,-0.1*\tikzbarheight) node[midway] {1 \\ P6};
			\draw[oddstyle] (27,-0.05*\tikzbarheight) rectangle (29.6,-0.1*\tikzbarheight) node[midway] {2 \\ P6};
			\draw[oddstyle] (29.6,-0.05*\tikzbarheight) rectangle (32.2,-0.1*\tikzbarheight) node[midway] {3 \\ P6};
			\draw[evenstyle] (27,0*\tikzbarheight) rectangle (29.7,-0.05*\tikzbarheight) node[midway] {1 \\ P7};
			\draw[evenstyle] (29.7,0*\tikzbarheight) rectangle (32.4,-0.05*\tikzbarheight) node[midway] {2 \\ P7};
			\draw[evenstyle] (32.4,0*\tikzbarheight) rectangle (35.1,-0.05*\tikzbarheight) node[midway] {3 \\ P7};
			\draw[evenstyle] (32.2,-0.05*\tikzbarheight) rectangle (34.7,-0.1*\tikzbarheight) node[midway] {1 \\ P8};
			\draw[evenstyle] (34.7,-0.05*\tikzbarheight) rectangle (37.2,-0.1*\tikzbarheight) node[midway] {2 \\ P8};
			\draw[evenstyle] (37.2,-0.05*\tikzbarheight) rectangle (39.7,-0.1*\tikzbarheight) node[midway] {3 \\ P8};
			\draw[oddstyle] (35.1,0*\tikzbarheight) rectangle (39.1,-0.05*\tikzbarheight) node[midway] {1 \\ P9};
			\draw[oddstyle] (39.1,0*\tikzbarheight) rectangle (43.1,-0.05*\tikzbarheight) node[midway] {2 \\ P9};
			\draw[oddstyle] (43.1,0*\tikzbarheight) rectangle (47.1,-0.05*\tikzbarheight) node[midway] {3 \\ P9};
			\draw[oddstyle] (39.7,-0.05*\tikzbarheight) rectangle (43.7,-0.1*\tikzbarheight) node[midway] {1 \\ P10};
			\draw[oddstyle] (43.7,-0.05*\tikzbarheight) rectangle (47.7,-0.1*\tikzbarheight) node[midway] {2 \\ P10};
			\draw[oddstyle] (47.7,-0.05*\tikzbarheight) rectangle (51.7,-0.1*\tikzbarheight) node[midway] {3 \\ P10};
			\draw [decorate,decoration={brace,amplitude=5pt,mirror,raise=4pt}] (47.7,-0.1*\tikzbarheight) -- (51.7,-0.1*\tikzbarheight) node [midway,yshift=-0.5cm] {forced-idle time};

        \end{tikzpicture}
    \label{fig_d}}

    \caption{
    Four example solutions to the instance of the Harmony scheduling problem that arises by applying the reduction to an instance $(a_1,a_2,a_3) = (6,2,4)$ ($n=3$) of Partition.
    We used $A=10$ for the illustration.
    On each GPU, every other pack is drawn with a blue pattern.
    A rectangle with label ``$b$ P$j$'' denotes the $j$-th pack being processed for microbatch $b$.
    }
    \label{fig:problem}
\end{figure*}

    \textbf{($\Longrightarrow$)}
    Let $X \subseteq \{1,...,n\}$ be a feasible solution to the Partition instance, i.e., $\sum_{i \in X} a_i = \sum_{i \not \in X} a_i$.
    We produce a packing as explained above: the first two and the last two layers are placed in singleton packs, and for each $i=1,...,n$, if $i \in X$, we form packs $\{3i,3i+1\}, \{3i+2\}$, and otherwise we form packs $\{3i\}, \{3i+1,3i+2\}$. Note that the pack containing $3i$ will be processed on GPU 1, and the pack containing $3i+2$ will be processed on GPU 2.
    See Fig.~\ref{fig_a} for an example.
    
    Now we will argue that GPUs are idle only during forced-idle times.
    Because $X$ was a feasible Partition solution, the total loads on both GPUs were equal.
    Therefore it is enough to show that neither GPU ever needs to wait for the other GPU
    (except for the forced-idle time at the beginning).
    (See Fig.~\ref{fig_c} for a situation where GPU 2 needed to wait for GPU 1 before starting pack 10 on microbatch 1.)
    This follows by our construction and because $A$ is large -- this is best seen from Fig.~\ref{fig_a}.
    More formally,
    consider the offset $O_j$ between the starting times of microbatch $1$ on pack $j+1$ and on pack $j$.
    We have $O_3 = 8A$,
    and one can prove by induction
    that $|O_j - 8A| \le B \cdot \sum_{i=1}^{(j-3)/2} a_i \le 3 \cdot \sum_{i=1}^n a_i \le A/2$
    for $j=3,5,...,2n+3$.
    Consequently, also for even $j=4,...,2n+2$ we have
    $|O_j - (B \cdot 5A - 8A)| = |O_j - 7A| \le A/2$.
    As all offsets are always between $6.5A$ and $8.5A$
    and pack lengths (except the first two and the last two)
    are between $4.5A$ and $5.5A$ (bounding $a_i < A/2$),
    the first microbatch of every pack
    begins processing while the second microbatch of the previous pack is being processed on the other GPU.
    Finally, as the loads are equal, at the end we have $O_{2n+3} = 8A$.
    
    \textbf{($\Longleftarrow$)}
    As explained above, any feasible solution to the \sysname scheduling instance puts the first two and the last two layers in singleton packs, and for $i=1,...,n$, puts layer $3i+1$ in a pack with $3i$, with $3i+2$, or alone. If $3i+1$ is put alone for any $i$, consider the smallest such $i$; at that point, an unforced idle time appears, which is best seen in Fig.~\ref{fig_b} and can be formally proved by arguing about offsets as above. In that case, makespan $T$ cannot be attained. So assume that layers $3i+1$ are never put alone. Let $X = \{ i=1,...,n : \{3i,3i+1\} \text{ is a pack}\}$.
    The total load on GPU 1 is $3 \cdot \left(8A + n \cdot 5A + \sum_{i \in X} a_i + 8A\right)$,
    and for GPU 2 we have the same but with $i \not \in X$ in place of $i \in X$.
    As $X$ is not a feasible solution to the Partition instance (none exists), these total loads are different and thus some GPU will have unforced-idle times. See Figs.~\ref{fig_c} and~\ref{fig_d} for two examples.
\end{proof}

\clearpage
\section{Additional Swap Analysis}
In this section, we provide additional swap load analysis that was not included in the main paper.
Beyond weights, we further analyze the swap volume for every swapped tensor in Table~\ref{tab:analysis}.
Weight gradients ($dW$) share a similar swap volume as $W$ for DP/PP with per-GPU memory virtualization but with $2m$ deducted from the forward pass, while \sysname{} DP/PP completely removes $dW$ swaps by generating $dW$ on GPU and consuming it right away with \textit{jit-update}.
For optimizer state $K$, the swap-in and -out volume is simply 2$\times$ per GPU during the update phase but with $N$ terms removed for all PP approaches.
For input/outputs $X,Y,dY,dX$, the swap does not take place for most approaches due to either \textit{p2p swaps} or the same microbatch running through all layers.
For stashed input $sX$, the swap volume is tied for DP/PP as they always require a CPU-GPU swap during the gap between forward pass's generation and backward pass's consumption.
From Table~\ref{tab:analysis}, we observe that \sysname{} PP offers the least swap volume for all tensors compared to all approaches.
We empirically show the advantages of \sysname{} in \secref{sec:eval}.
\begin{table}[h]
\caption{Detailed analytical comparison of swap volume for every tensor. $N$ GPUs and $m$ microbatches per GPU are used to compute the swap volume of an entire minibatch.}
\label{tab:analysis}
\centering
\vspace{-2ex}
\begin{tabular}{c}
\includegraphics[width=0.98\linewidth]{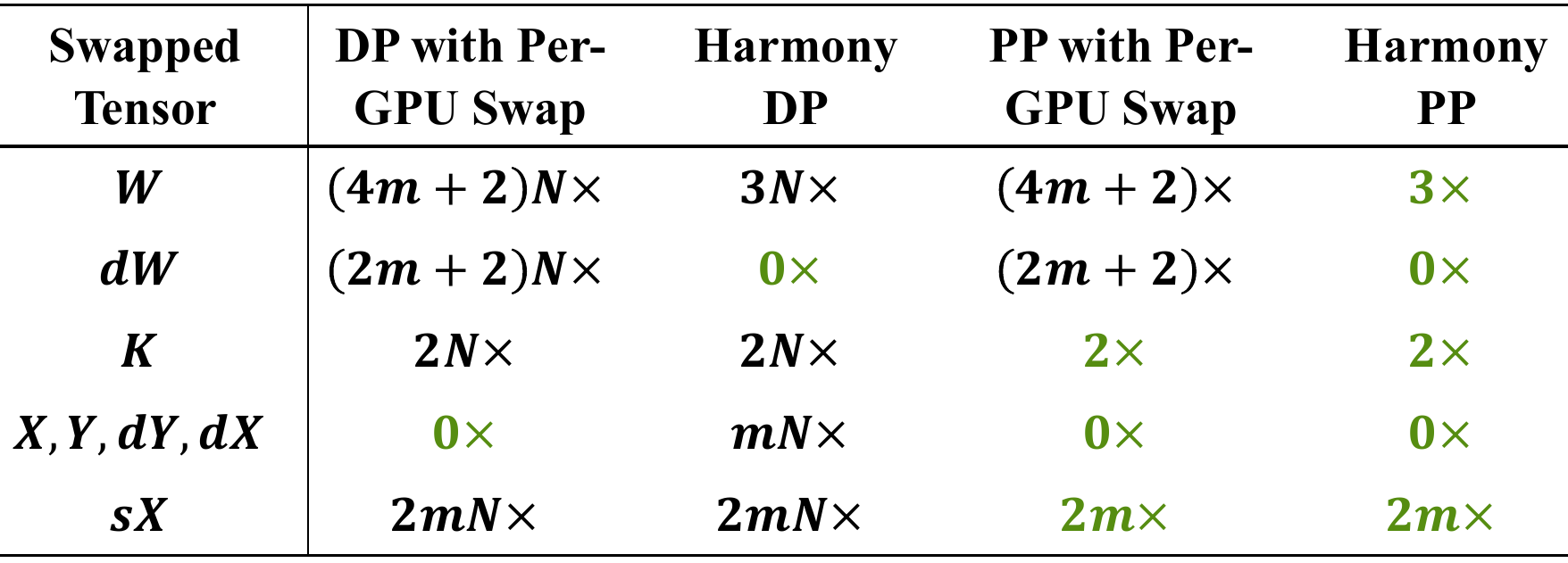}
\end{tabular}
\end{table}

\clearpage
\section{Additional Scheduler Algorithms}
In this section, we provide additional algorithms used by \sysname{}'s Scheduler that were not included in the main paper.

\niparagraph{Task Graph Generation.} 
Algorithm~\ref{alg:tgraph_gen} shows how to generate a task graph from a given configuration. 
This is a simplified view, showing only \sysname{} PP, only input/output for forward tasks, and concise channel setup for weights. 
Note that each layer has only one channel, e.g., input activation ``X'', enforced by \sysname{}'s Decomposer that consolidates multiple tensors into one input (including serializing branches by relaying tensors in a sequential graph to their destination layer).
\begin{algorithm}[!t]
\small
\DontPrintSemicolon
\newcommand\mycommfont[1]{\footnotesize\ttfamily\textcolor{blue!50!black}{#1}}
\SetCommentSty{mycommfont}
\SetKwInOut{Input}{Input}
\SetKwInOut{Output}{Output}
\caption{Task Graph Generation $\rho$ (Simplified)}
\label{alg:tgraph_gen}
\Input{forward/backward microbatch size $U_F$/$U_B$,\\
       forward/backward layer packs $P_F$/$P_B$,\\
       number of GPUs $N$,\\
       minibatch size $D$}
\Output{generated task graph $G$}
$G \gets []$\tcp*[l]{empty list}
$idx \gets 0$\tcp*[l]{task index}
\tcp{create group of microbatch sizes}
$u_F \gets [U_F^{(1)},U_F^{(2)},\ldots,U_F^{(D/U_F)}]$\;
$u_B \gets [U_B^{(1)},U_B^{(2)},\ldots,U_B^{(D/U_B)}]$\;
\tcp{create forward tasks}
\For{$p \gets P_F[0]$ \textbf{to} $P_F[end]$}
{
    $t \gets Task()$\tcp*[l]{init a task}
    $t.i \gets idx$++\tcp*[l]{task's index}
    $t.p \gets p$\tcp*[l]{task's layer pack}
    $t.\tau \gets F$\tcp*[l]{task's type}
    $t.u \gets u_F$\tcp*[l]{task's group of microbatch sizes}
    $G.Append(t)$\;
}
\tcp{create backward and update tasks}
\For{$p \gets P_B[end]$ \textbf{to} $P_B[0]$}
{
    $t \gets Task(idx\text{++},\ p,\ B,\ u_B)$\tcp*[l]{a backward task}
    $G.Append(t)$\;
    $t \gets Task(idx\text{++},\ p,\ A)$\tcp*[l]{a jit update task}
    $G.Append(t)$\;
}
\tcp{bind tasks to devices for a round-robin pipeline}
\For{$t \gets G[0]$ \textbf{to} $G[end]$}
{
    \uIf{$t.\tau$ is $F$}
    {
        $t.d \gets GPU\#(t.i \bmod N)$\tcp*[l]{bind to GPU-ID}
    }
    \uElseIf{$t.\tau$ is $B$}
    {
        $t.d \gets GPU\#((|P_F|+\frac{t.i-|P_F|}{2}) \bmod N)$\;
    }
    \Else
    {
        $t.d \gets CPU\#((|P_F|+\frac{t.i-1-|P_F|}{2}) \bmod N)$\;
    }
}
\tcp{init tasks' input and output based on Figure~\ref{fig:analysis}a}
\For{$t \gets G[0]$ \textbf{to} $G[end]$}
{
    $t.in \gets \{\}$\tcp*[l]{dictionary}
    $t.out \gets \{\}$
}
\tcp{set input/output dependency with runtime channels}
\For{$t \gets G[0]$ \textbf{to} $G[end]$}
{
    \uIf{$t.\tau$ is $F$}
    {
        $c \gets Channel(P2P,\ SrcTask=t.i-1)$\;
        $t.in[X] \gets \{\ t.p[0] : c\ \}$\;
        $t.in[W] \gets \{\ t.p[0] : Channel(SharedMem),$\;
        $\quad\quad\quad\quad\quad t.p[1] : Channel(PinnedGPU),$\;
        $\quad\quad\quad\quad\quad \ldots\ \}$\tcp*[l]{one channel per layer}
        $c \gets Channel(P2P,\ DstTask=t.i+1)$\;
        $t.out[Y] \gets \{\ t.p[end] : c\ \}$\;
        \tcp{find stash activations for backward tasks}
        $G_B = Filter(G,\ ``\tau\ is\ B")$\;
        $G_B' = Filter(G_B,\ ``first\ layer\ in\ current\ t.p")$\;
        \For{$t' \gets G_B'[0]$ \textbf{to} $G_B'[end]$}
        {
            $c \gets Channel(Message,\ DstTask=t'.i)$\;
            $t.out[X] \gets \{\ t'.p[0] : c\ \}$\;
        }
    }
    \uElseIf{$t.\tau$ is $B$}
    {
        $\ldots$\;
    }
    \Else
    {
        $\ldots$\;
    }
}
\KwRet{$G$}
\end{algorithm}

\clearpage
\section{Additional Evaluation Results}
In this section, we provide additional experimental results that were not included in the main paper.

\begin{figure}[!h]
  \vspace{-2ex}
  \centering
  \includegraphics[width=\linewidth]{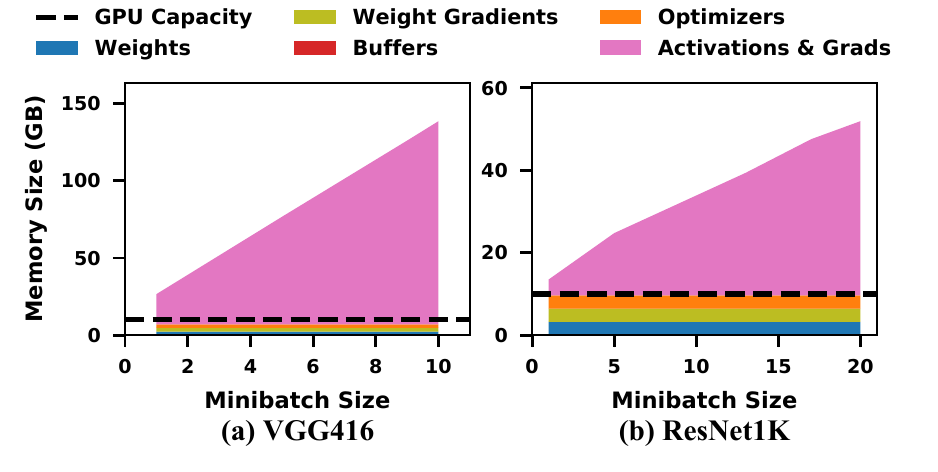}
  \vspace{-6ex}
  \caption{Memory footprint statistics for training CNNs at different minibatch sizes (using virtualized GPU memory).}
  \label{fig:memory_scale_vggresnet}
\end{figure}

\niparagraph{Memory Footprint.}
Figure~\ref{fig:memory_scale_vggresnet} shows the memory requirements for training large CNNs across different minibatch sizes, when GPU memory is virtualized for a single GPU.

\begin{figure}[!h]
\vspace{-2ex}
\centering
    \begin{subfigure}[b]{0.48\textwidth}
      \includegraphics[width=1\linewidth]{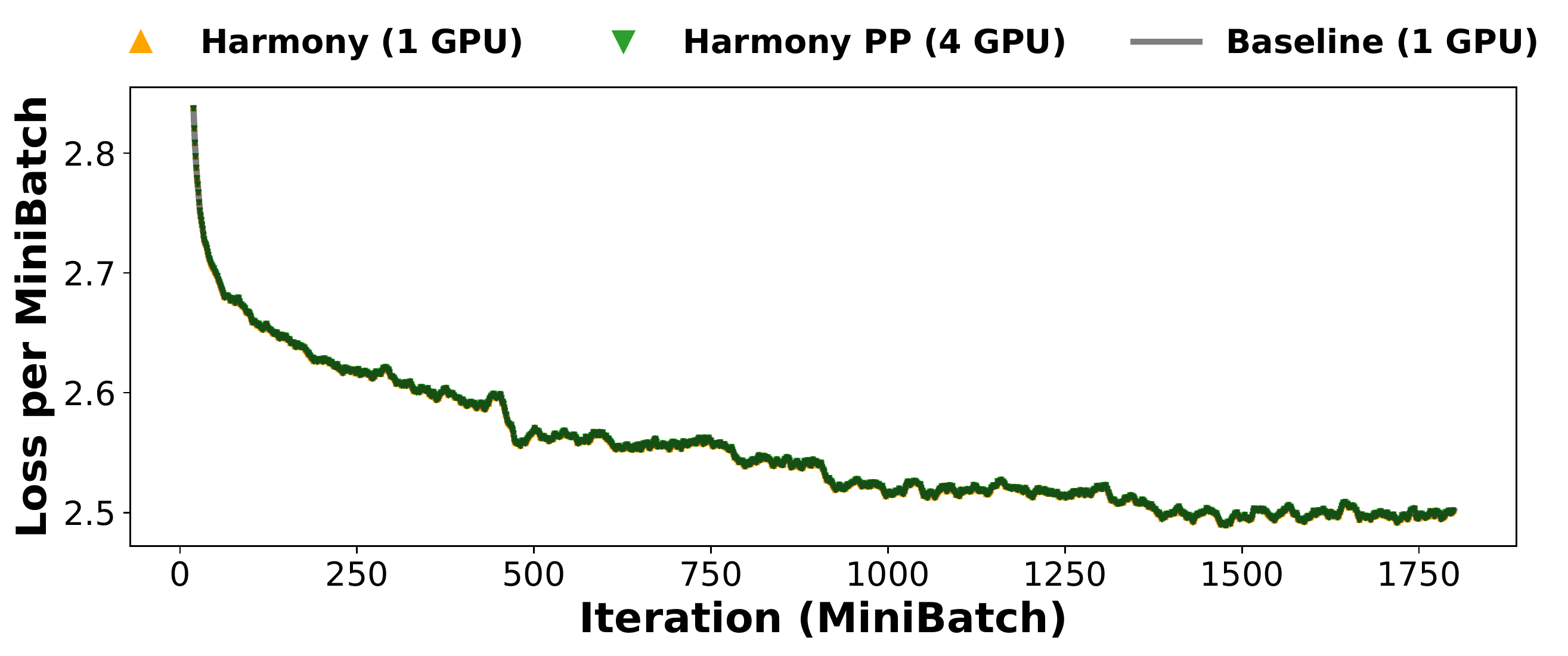}
      \vspace{-4ex}
      \caption{\sysname vs. single-GPU baseline.}
    \end{subfigure}
    \vspace{1ex}
    \begin{subfigure}[b]{0.47\textwidth}
      \centering
      \includegraphics[width=1\linewidth]{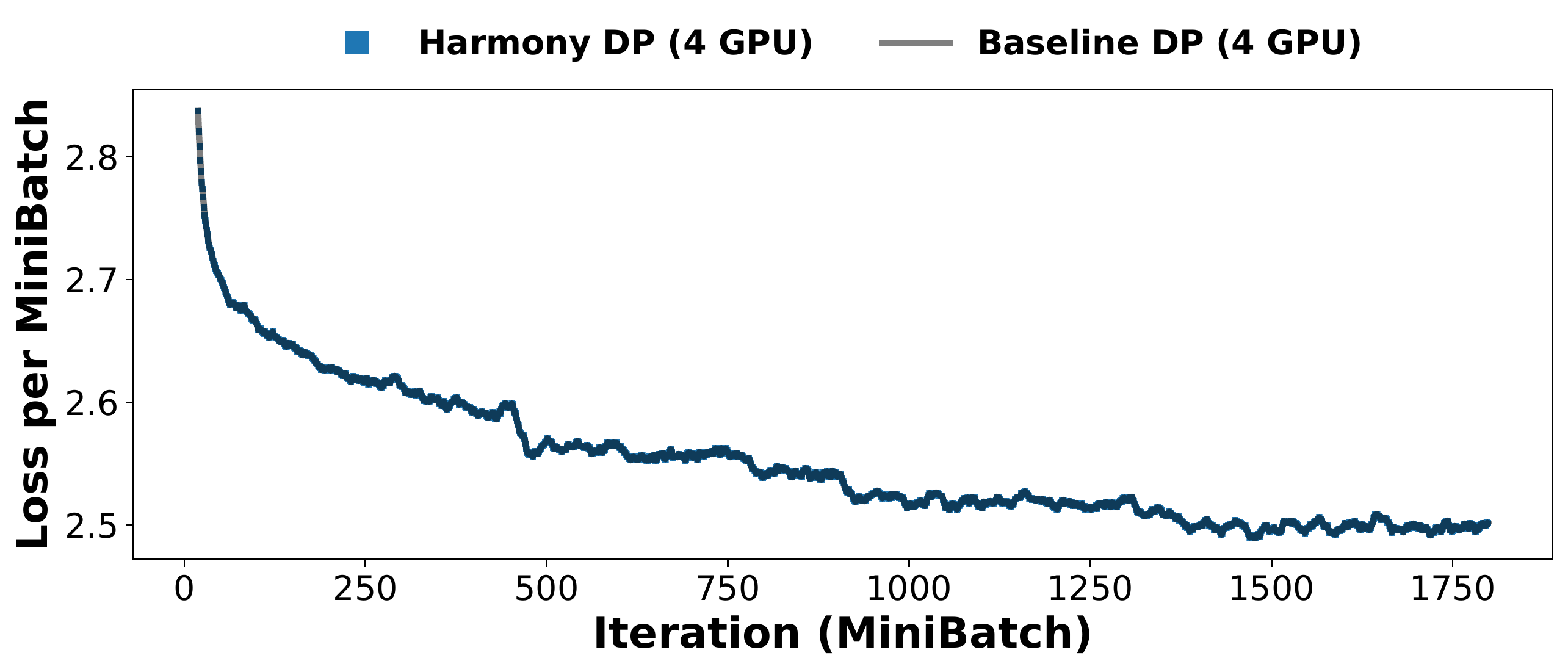}
      \vspace{-4ex}
      \caption{\sysname vs. data-parallelism baseline.} 
    \end{subfigure}
\vspace{-3ex}
\caption{Correctness of \sysname. An example of fine-tuning GPT2-Medium (0.3B) on WikiText103 with reported hyper-parameters~\cite{BlogGPT2} and baseline code~\cite{PTGPT2}. 
\sysname matches the baseline exactly for every minibatch.}
\label{fig:curve_gpt2}
\end{figure}

\niparagraph{Correctness of Training in \sysname{}.}
Figure~\ref{fig:curve_gpt2} compares the training curve of \sysname{} with baselines for fine-tuning GPT2-Medium under the same settings. 
The exact match validates that \sysname offers synchronous SGD convergence as baselines.
The same results are also observed in evaluation accuracy in Table~\ref{tab:fine_tuning}.
\begin{table}[!h]
\centering
\footnotesize
\caption{Correctness of \sysname. Comparison of evaluation accuracy/perplexity of fine-tuning BERT-Large on MRPC and GPT2-Medium on WikiText103. \sysname shares the same setting (hardware, hyper-parameters, and code) as the baseline~\cite{devlin2018bert,PTBERT,BlogGPT2,PTGPT2} and matches baseline accuracy.}
\label{tab:fine_tuning}
\vspace{-2ex}
\begin{tabular}{@{}c|ccc|cc@{}}
Model & Baseline   & \textbf{Harmony}    & \textbf{Harmony PP}      & Baseline DP & \textbf{Harmony DP}\\
      & (1 GPU)    & (1 GPU)             & (4 GPU)                  & (4 GPU)     & (4 GPU)\\ 
\hline
BERT  & 88.0\%              & 88.0\%                      & 88.0\%  & 87.3\%      & 87.3\%\\
GPT2  & 12.1                & 12.1                        & 12.1    & 12.1        & 12.1
\end{tabular}
\end{table}

\niparagraph{Equal Configuration vs. Distinct Configuration.}
Table~\ref{tab:equi_fb} compares the training time between \textit{Equi-FB} (the same configuration for forward and backward pass) and \textit{Distinct-FB} (different configuration for forward and backward pass) for configurations search in \sysname.
We observe that:
1) Separating forward and backward configuration (i.e., \textit{Distinct-FB}) enables a richer design space, thus offering a higher performance (up to 29.1\%),
and 2) Convolutional neural networks benefit more from \textit{Distinct-FB} than Transformers, due to the irregularity of per-layer characteristics (compute time, memory footprint, activation size) in CNNs.
\begin{table}[!h]
\vspace{-2ex}
\small
\caption{Comparison of different packing strategies for \sysname{}'s configuration search (with a minibatch size of 16).  Iteration times are measured for deployed training runs.}
\label{tab:equi_fb}
\vspace{-2ex}
\begin{tabular}{@{}cccc@{}}
\toprule
Model    & Equi-FB (sec) & Distinct-FB (sec) & Improvement     \\ \midrule
BERT96   & 19.438        & 18.893            & 2.8\%           \\
GPT2     & 44.700        & 43.924            & 1.7\%           \\
VGG416   & 10.289        & 9.027             & \textbf{12.3\%} \\
ResNet1K & 5.456         & 3.868             & \textbf{29.1\%} \\ \bottomrule
\end{tabular}
\end{table}

\niparagraph{Layer Packs.}
Table~\ref{tab:detail_packs} shows the details of layer packs ($P_F$, $P_B$) for different models evaluated in Table~\ref{tab:search_time} of the main paper. 
\begin{table}[h!]
\vspace{-2ex}
\footnotesize
\center
\caption{Detailed scheduling result of \sysname PP with 4 GPUs, minibatch size 64, and balanced time packing.}
\label{tab:detail_packs}
\vspace{-2ex}
\begin{tabular}{@{}lll@{}}
\toprule
Model                     & \multicolumn{2}{l}{Layer Packs}                                                                                                                                                                                                                                       \\ \midrule
\multirow{2}{*}[-2.0em]{BERT96}   & $P_F$ & \begin{tabular}[c]{@{}l@{}}L0-3, L4-7, L8-11, L12-15, L16-19, \\ L20-22, L23-26, L27-30, L31-34, L35-38, \\ L39-42, L43-46, L47-49, L50-53, L54-57, \\ L58-61, L62-65, L66-69, L70-72, L73-76, \\ L77-80, L81-84, L85-88, L89-92\end{tabular}                                                                                                                                       \\ \cmidrule(l){2-3} 
                          & $P_B$ & \begin{tabular}[c]{@{}l@{}}L0-3, L4-7, L8-11, L12-15, L16-19, \\ L20-22, L23-26, L27-30, L31-34, L35-38, \\ L39-42, L43-46, L47-49, L50-53, L54-57, \\ L58-61, L62-65, L66-69, L70-72, L73-76, \\ L77-80, L81-84, L85-88, L89-92, L93-99\end{tabular} \\ \midrule
\multirow{2}{*}[-2.0em]{GPT2}     & $P_F$ & \begin{tabular}[c]{@{}l@{}}L0-4, L5-9, L10-14, L15-18, L19-23, \\ L24-28, L29-32, L33-37, L38-42, L43-47\end{tabular}                                                                                                                                          \\ \cmidrule(l){2-3} 
                          & $P_B$ & \begin{tabular}[c]{@{}l@{}}L0-2, L3-5, L6-8, L9-11, L12-14, \\ L15-17, L18-20, L21-23, L24-26, L27-29, \\ L30-32, L33-35, L36-38, L39-41, L42-44, \\ L45-47, L48-51\end{tabular} \\ \midrule
\multirow{2}{*}[-2em]{VGG416}   & $P_F$ & \begin{tabular}[c]{@{}l@{}}L0-14, L15-30, L31-45, L46-61, L62-76, \\ L77-98, L99-122, L123-146, L147-171, \\ L172-199, L200-228, L229-256, L257-284, \\ L285-312, L313-353\end{tabular}                                                                                                                      \\ \cmidrule(l){2-3} 
                          & $P_B$ & \begin{tabular}[c]{@{}l@{}}L0-14, L15-30, L31-45, L46-61, L62-76, \\ L77-98, L99-122, L123-146, L147-171, \\ L172-199, L200-228, L229-256, L257-284, \\ L285-312, L313-353, L354-416\end{tabular}                                                             \\ \midrule
\multirow{2}{*}[-1.5em]{ResNet1K} & $P_F$ & \begin{tabular}[c]{@{}l@{}}L0-429, L430-887\end{tabular}                                                                                                                                                           \\ \cmidrule(l){2-3} 
                          & $P_B$ & \begin{tabular}[c]{@{}l@{}}L0-93, L94-187, L188-293, L294-427, \\ L428-551, L552-653, L654-755, L756-887, \\ L888-1029\end{tabular}                             \\ \bottomrule
\end{tabular}
\end{table}

\begin{figure}[h!]
\vspace{-2ex}
\centering
    \begin{subfigure}[b]{0.48\textwidth}
      \centering
      \includegraphics[width=\linewidth]{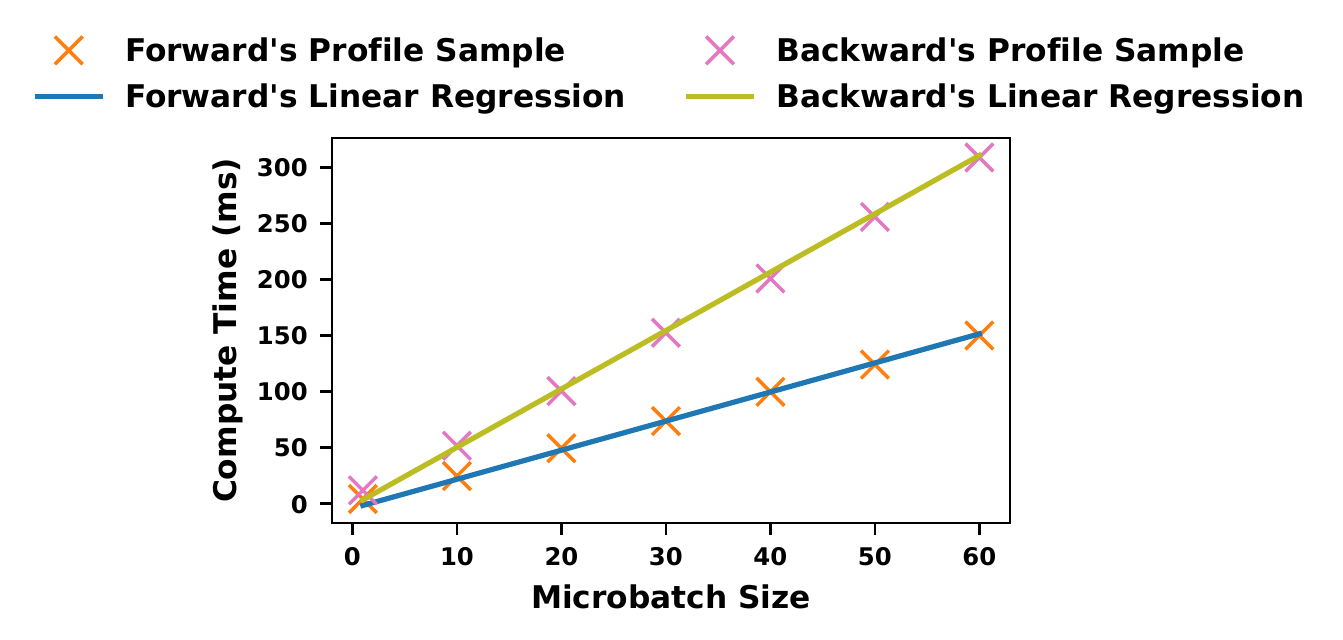}
      \vspace{-5ex}
      \caption{Estimating layer compute times across microbatch sizes with \textit{linear regression}.} 
      \label{fig:linear_regression_short}
    \end{subfigure}
    
    \begin{subfigure}[b]{0.48\textwidth}
      \centering
      \vspace{1em}
      \includegraphics[width=\linewidth]{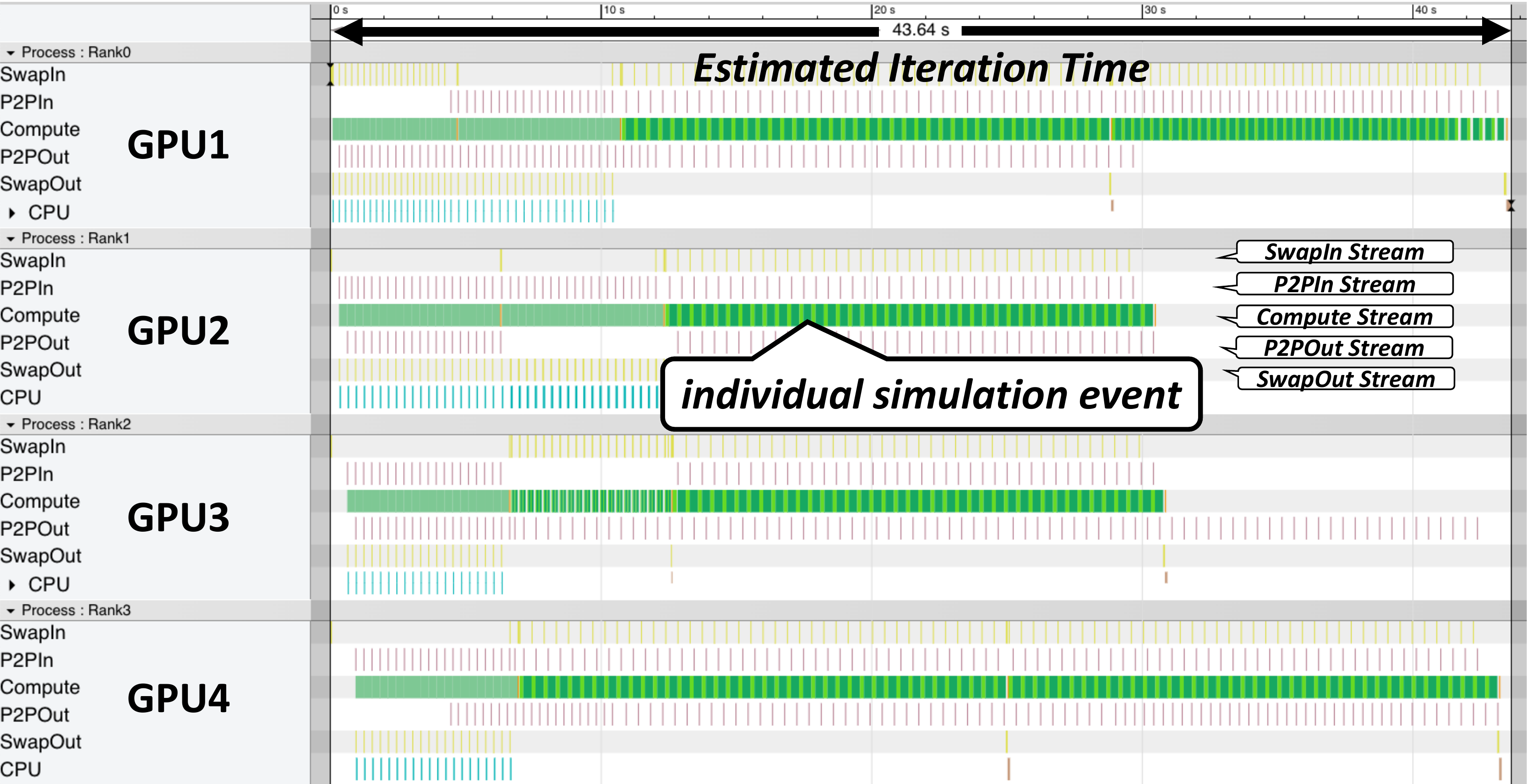}
      \caption{Estimating training time with an \textit{event-driven simulation}.}
      \label{fig:sim_view_short}
    \end{subfigure}
\vspace{-4ex}
\caption{The two kinds of ``estimations'' in \sysname{}. Figure~\ref{fig:sim_view_short} corresponds to Figure~\ref{fig:est_acc}.} 
\label{fig:two_costs_short}
\end{figure}

\niparagraph{Profiling and Simulation Details.}
\textit{Estimated profiles} (Figure~\ref{fig:two_costs_short}(a)) should not be confused with \textit{estimated end-to-end training iteration time} (Figure~\ref{fig:two_costs_short}(b) and Figure~\ref{fig:est_acc}). 
The \textit{profiles} capture the characterization of each layer with a \textit{linear regression} for interpolation across a wide range of microbatch sizes that our Profiler doesn't sample (\secref{sec:profiler}).
Without regression, one can risk brute-force profiling every possible microbatch size (e.g., from 1 to 8196) for every layer (e.g., 1024 layers) over hundreds of profiling iterations, which in practice takes more time than training the model itself.
Linear regression offers low overhead and results in good accuracy as shown in Figure~\ref{fig:linear_regression_short}.
By contrast, \textit{estimated training time} is obtained from an \textit{event-driven simulator} (\secref{sec:config_search}), which takes a \sysname{} task graph and the aforementioned profiles for estimating end-to-end iteration time, as shown in Figure~\ref{fig:sim_view_short}.

\begin{figure*}[h]
  \centering
  \includegraphics[width=0.95\textwidth]{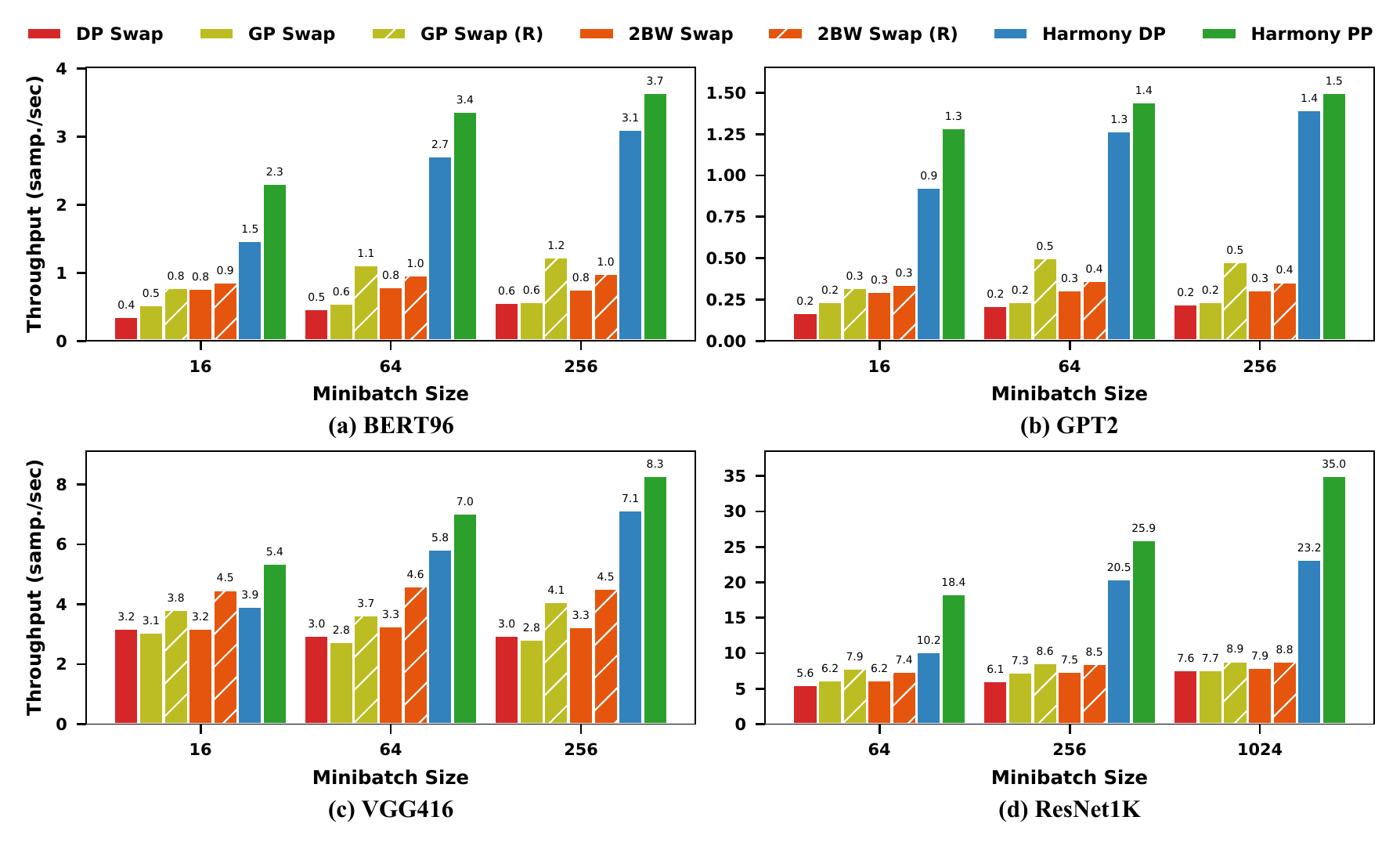}
  \vspace{-3ex}
  \caption{Performance comparison with per-GPU swap baselines by training different models with various minibatch sizes on 4 GPUs. Each group of bars represents one minibatch size. \textit{R} denotes the usage of recompute for activations.}
  \label{fig:throughput_vs_baselines_raw}
  \vspace{-2ex}
\end{figure*}

\begin{figure*}[h]
  \centering
  \includegraphics[width=0.95\linewidth]{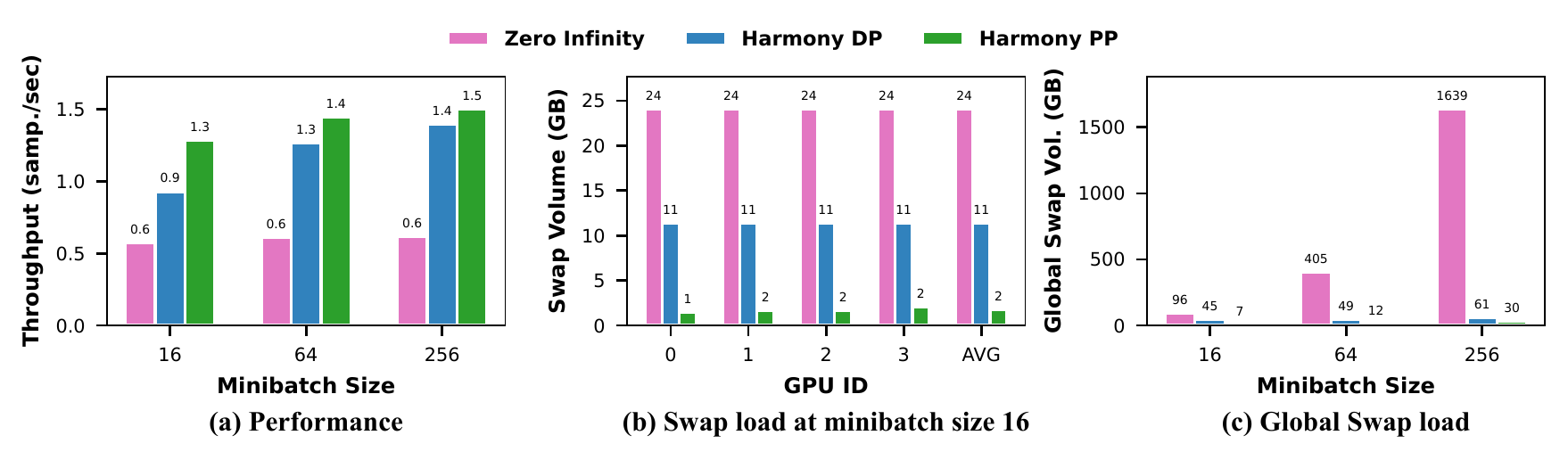}
  \vspace{-4ex}
  \caption{Comparison with \zinf for training GPT2 (1.5B) on 4 GPUs. CPU-GPU swap volume is measured per minibatch.} 
  \label{fig:gpt2_vs_zeroinf_raw}
  \vspace{-2ex}
\end{figure*}

\begin{figure}[h]
  \centering
  \includegraphics[width=\linewidth]{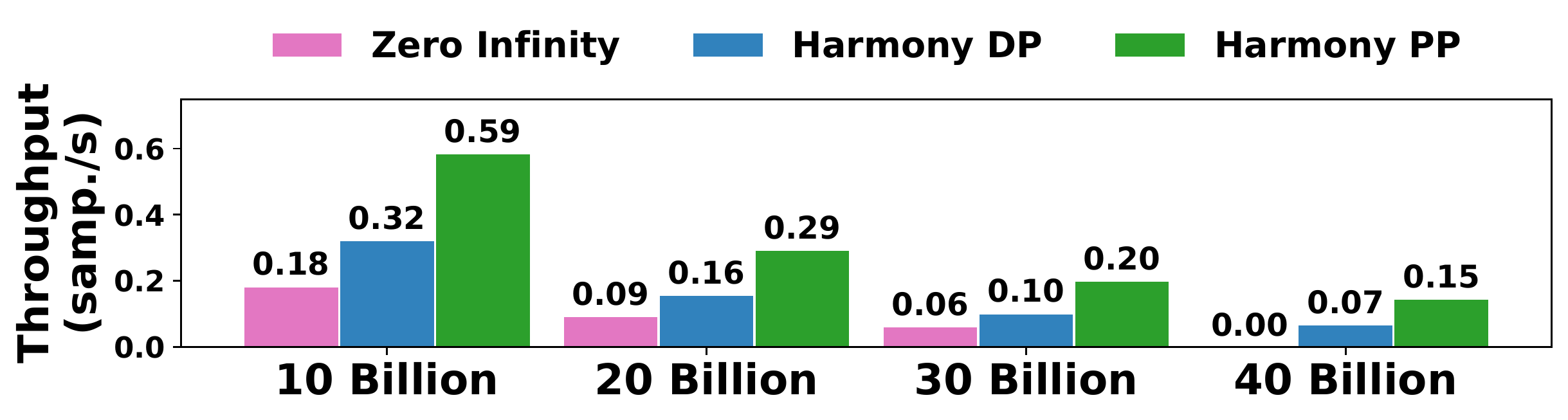}
  \vspace{-5ex}
  \caption{Training massive models of 10s billions of parameters at the limit of single-server CPU memory capacity (750GB CPU and eight GTX1080Tis).} 
  \label{fig:two_numa_largest_models_raw}
  \vspace{-1ex}
\end{figure}

\begin{figure}[h]
  \centering
  \includegraphics[width=\linewidth]{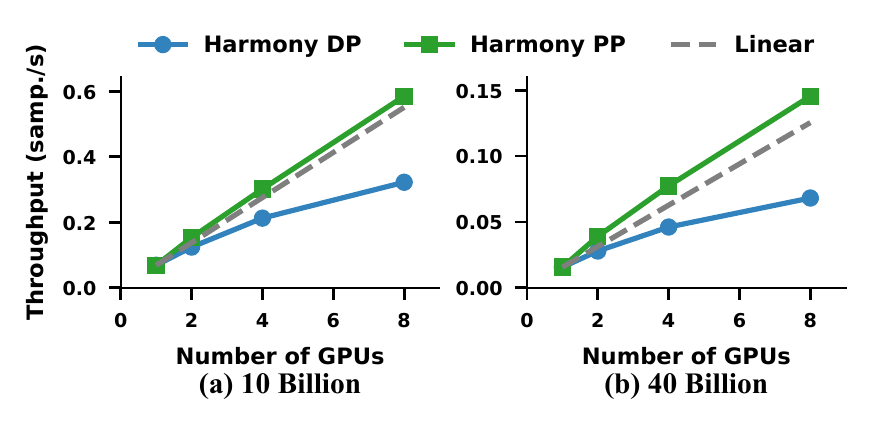}
  \vspace{-6ex}
  \caption{Scalability of \sysname in training massive models of 10s of billions of parameters with eight GTX1080Tis.} 
  \label{fig:scalability_largest_models_onecol_raw}
\end{figure}

\begin{figure*}[h]
  \centering
  \includegraphics[width=0.9\linewidth]{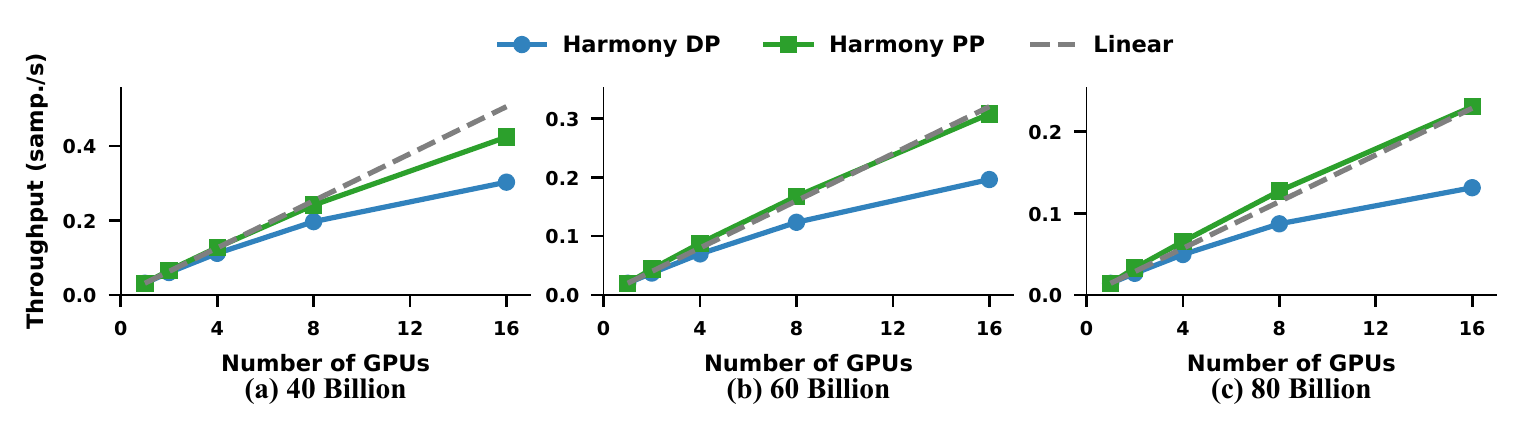}
  \vspace{-4ex}
  \caption{Scalability of \sysname in training massive models of 10s of billions of parameters with 16 V100s. The 80-Billion model saturates the CPU memory capacity (1.5TB).} 
  \label{fig:scalability_dgx_raw}
\end{figure*}

\niparagraph{Raw Throughput.}
Figures~\ref{fig:throughput_vs_baselines_raw}, \ref{fig:gpt2_vs_zeroinf_raw}, \ref{fig:two_numa_largest_models_raw}, \ref{fig:scalability_largest_models_onecol_raw}, \ref{fig:scalability_dgx_raw} show the raw throughput numbers corresponding to their normalized counterparts in \secref{sec:eval} of the main paper. 

\clearpage

\end{document}